\pdfoutput=1
\documentclass{aa}  
\usepackage{graphicx,natbib}
\usepackage{caption}
\usepackage{subcaption}

\usepackage{txfonts}
\def\kms{km~s$^{-1}$}

\def\my{\mbox{$M_{\odot}$~yr$^{-1}$}}

\newcommand{\nc}{\newcommand}

\nc{\RAJ}[4]{$\alpha(J2000) = {#1}^{\rm h}{#2}^{\rm m}{#3}\fs{#4}$}
\nc{\DecJ}[4]{$\delta(J2000) = {#1}\degr {#2}\arcmin {#3}\farcs{#4}$}

\begin{document}

   \title{Molecular line polarisation from the circumstellar envelopes of Asymptotic Giant Branch stars}

  \titlerunning{Molecular line polarisation in CSEs}


   \author{W.~H.~T. Vlemmings
          \inst{1}\fnmsep\thanks{wouter.vlemmings@chalmers.se}
          \and
          B. Lankhaar\inst{2,1}
          \and
          L. Velilla-Prieto\inst{3}
          }

   \institute{Department of Space, Earth and Environment, Chalmers University of Technology, 412 96, Gothenburg, Sweden
   \and
   Leiden Observatory, Leiden University, Post Office Box 9513, 2300 RA Leiden, Netherlands
   \and
   Department of Molecular Astrophysics, Institute of Fundamental Physics, Consejo Superior de Investigaciones Cient\'ificas, Serrano 123, 28006 Madrid, Spain}

   \date{v4 23-Mar-2024}

 
  \abstract
   {Polarisation observations of masers in the circumstellar envelopes (CSEs) around Asymptotic Giant Branch (AGB) stars have revealed strong magnetic fields. However, masers probe only specific lines-of-sight through the CSE. Non-masing molecular line polarisation observation can more directly reveal the large scale magnetic field morphology and hence probe the effect of the magnetic field on AGB mass-loss and the shaping of the AGB wind.}
   {Observations and models of CSE molecular line polarisation can now be used to describe the magnetic field morphology and estimate its strength throughout the entire CSE.}
   {We use observations taken with the Atacama Large Millimeter/submillimeter Array (ALMA) of molecular line polarisation in the envelope of two AGB stars (CW~Leo and R~Leo). We model the observations using the multi-dimensional polarised radiative transfer tool PORTAL.}
   {We find linearly polarised emission, with maximum fractional polarisation on the order of a few percent, in several molecular lines towards both stars.
   Towards R~Leo we also find a high level of linear polarisation (up to $\sim35\%$) for one of the SiO~$v=1$ maser transitions. We can explain the observed differences in polarisation structure between the different molecular lines by alignment of the molecules by a combination of the Goldreich-Kylafis effect and radiative alignment effects. We specifically show that the polarisation of CO traces the morphology of the magnetic field. Competition between the alignment mechanisms allows us to describe the behaviour of the magnetic field strength with radius throughout the circumstellar envelope of CW~Leo. The magnetic field strength derived using this method is inconsistent with the magnetic field strength derived using a structure function analysis of the CO polarisation and the strength previously derived using CN Zeeman observations. In contrast with CW~Leo, the magnetic field in the outer envelope of R~Leo appears to be advected outwards by the stellar wind.}
   {The ALMA observations and our polarised radiative transfer models show the power of using multiple molecular species to trace the magnetic field behaviour throughout circumstellar envelope. While the observations appear to confirm the existence of a large scale magnetic field, further observations and modelling is needed to understand the apparent inconsistency of the magnetic field strength derived with different methods in the envelope of CW~Leo.}


   \maketitle
%

\section{Introduction}

\begin{figure*}
     \centering
     \begin{subfigure}[b]{0.49\textwidth}
         \centering
         \includegraphics[width=\textwidth]{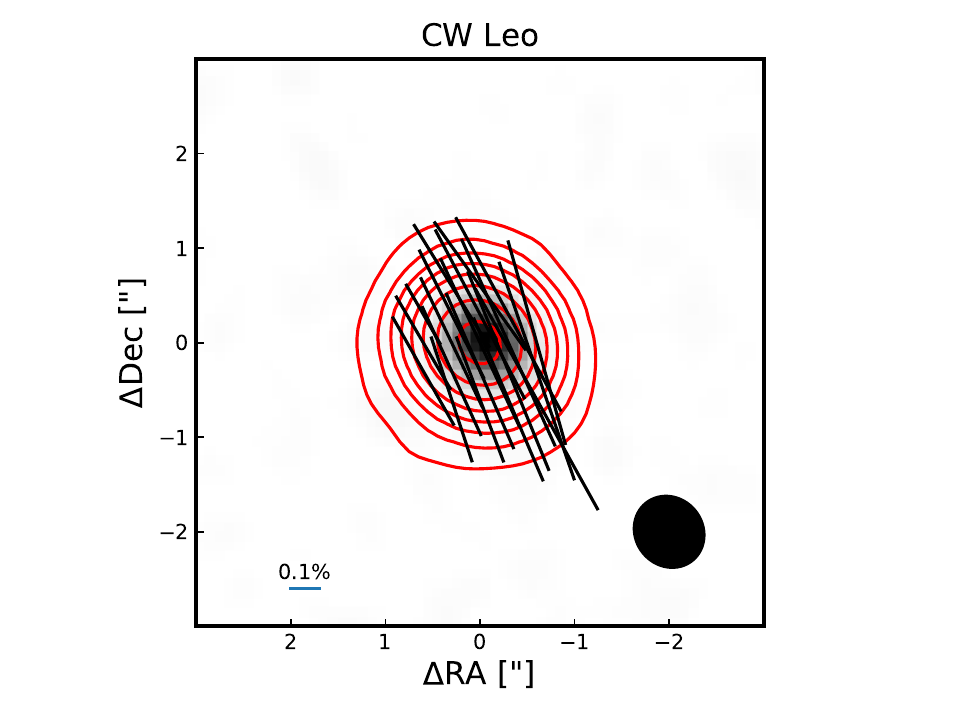}
     \end{subfigure}
     \hfill
     \begin{subfigure}[b]{0.49\textwidth}
         \centering
         \includegraphics[width=\textwidth]{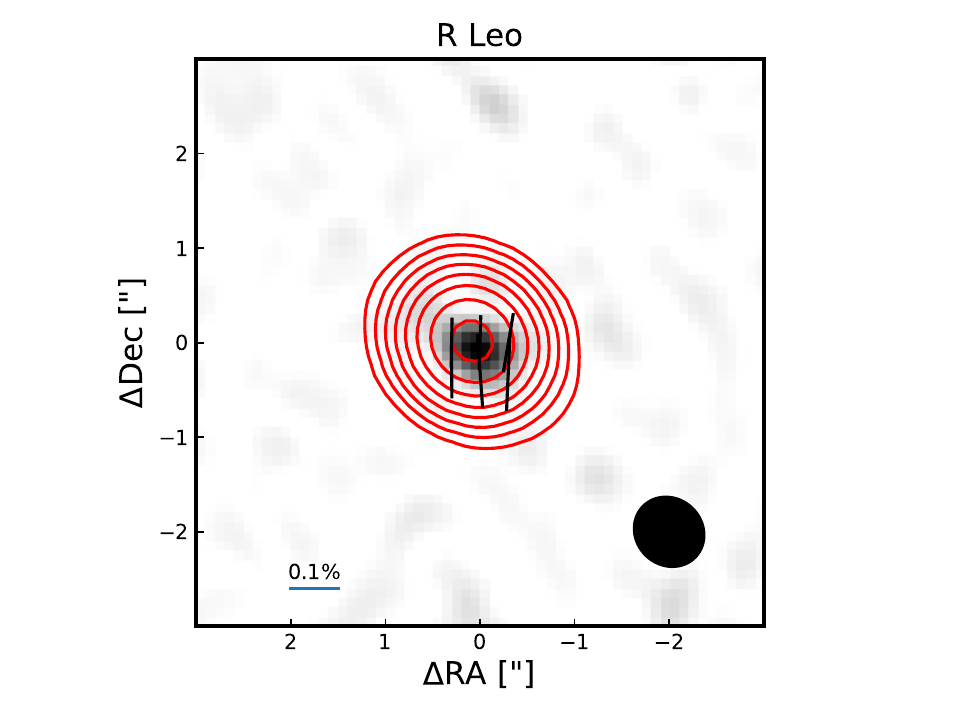}
     \end{subfigure}
     \hfill
        \caption{Polarised continuum emission (greyscale) and the 338~GHz continuum contours of CW~Leo (left) and R~Leo (right). The polarisation peaks at $6.9$, and  $0.4$~mJy~beam$^{-1}$ for CW~Leo and R~Leo respectively. The contours are drawn at $0.625, 1.25, 2.5, 5, 10, 20, 40,$ and $80\%$ of the peak emission ($1.059$, and $0.302$~Jy~beam$^{-1}$ for CW~Leo and R~Leo respectively). The line segments indicate the linear polarisation direction where polarised emission is detected at $>5\sigma_P$. The uncertainty on the direction is $\lesssim 6^\circ$. The segments are scaled as indicated by the horizontal bar. The filled ellipses indicates the beam size of the observations.}
        \label{fig:contpol}
\end{figure*}

Magnetic fields have been observed in the envelopes of many evolved asymptotic giant branch (AGB) stars \citep{Vlemmings19}. These stars, with
main-sequence masses between $1-8$~M$_\odot$, undergo significant mass-loss that is an important source of interstellar enrichment by nucleosynthesis elements and dust \citep{HO18}. The winds that cause this mass-loss are mainly driven by radiation pressure on the dust that forms within a few stellar radii of the stellar surface. As the surface magnetic field strength for AGB stars has been inferred to be of the order of several Gauss, magnetic fields can potentially play an important role in the initial ejection of material from the star as well as in shaping the circumstellar envelope. Additionally, at early stages after the AGB phase, magnetic fields are thought to play a role in launching collimated outflows that are important in the subsequent evolution from AGB to Planetary Nebula \citep[e.g.][]{Vlemmings06, PerezSanchez13}. Only for one AGB star, $\chi$~Cyg, the magnetic field strength, which is of the order of $2-3$~G, has been directly measured on the surface \citep{Lebre14}. In other cases, the magnetic field strength at the stellar surface is estimated by extrapolating, from the circumstellar envelope (CSE) to the surface, Zeeman measurements in compact SiO, OH and H$_2$O masers \citep[e.g.][]{Vlemmings02, Vlemmings05, Herpin06, LealFerreira13, Gonidakis14}, or in a single case from CN \citep{Duthu17}.  In order to perform the extrapolation, multiple maser species that are excited at different locations throughout the circumstellar envelope (CSE) can be used, but generally a standard dipole, toroidal, solar-type or radial magnetic field configuration is assumed. This leads to significant uncertainties, and as a result, the role of magnetic fields in supporting AGB mass-loss is still unclear.

Molecular line polarisation from non-maser lines can provide important further information, as it can constrain the morphology of the magnetic fields and improve the extrapolation of the measured field strength and also directly provide constraints on the field strength and its effect on the shape of the CSE. In what is known as the Goldreich-Kylafis (GK) effect \citep[e.g.][]{GK82}, even the presence of a weak magnetic field will lead to molecular line polarisation, as the magnetic sublevels of the involved rotational states are differently populated in an anisotropic radiation field. The GK-effect of the CO molecule has been observed in star forming regions \citep[e.g.][]{Cortes05, Beuther10, Li11} and a recent tentative detection has been made in proto-planetary discs \citep{Stephens20, Teague21}. Around (post-)AGB stars, the GK-effect in CO has been detected in five sources \citep{Girart12, Vlemmings12, Huang20, Vlemmings23}. The same observations revealed linear polarisation in other non-masing molecular lines, such as SiO, CS, and SiS. 

Around AGB stars, linear polarisation can also arise from molecular lines that originate from molecules with a preferred rotation axis caused by a strong radial infrared radiation field from the central star \citep{Morris85}.
Detailed modelling of the involved lines is needed to determine the mechanism that causes the observed polarisation. As the envelope, magnetic field and radiation field around evolved stars can be asymmetric, thus, multi-dimensional polarised radiative transfer is needed. In this paper we present observational and modelling constraints on the molecular line polarisation properties of two well-known AGB stars, CW\,Leo and R\,Leo. The stars were observed with ALMA and the modelling is done using the PORTAL polarised radiative transfer code \citep{Lankhaar20}.

\begin{figure*}[ht!]
\centering
\includegraphics[width=0.95\textwidth]{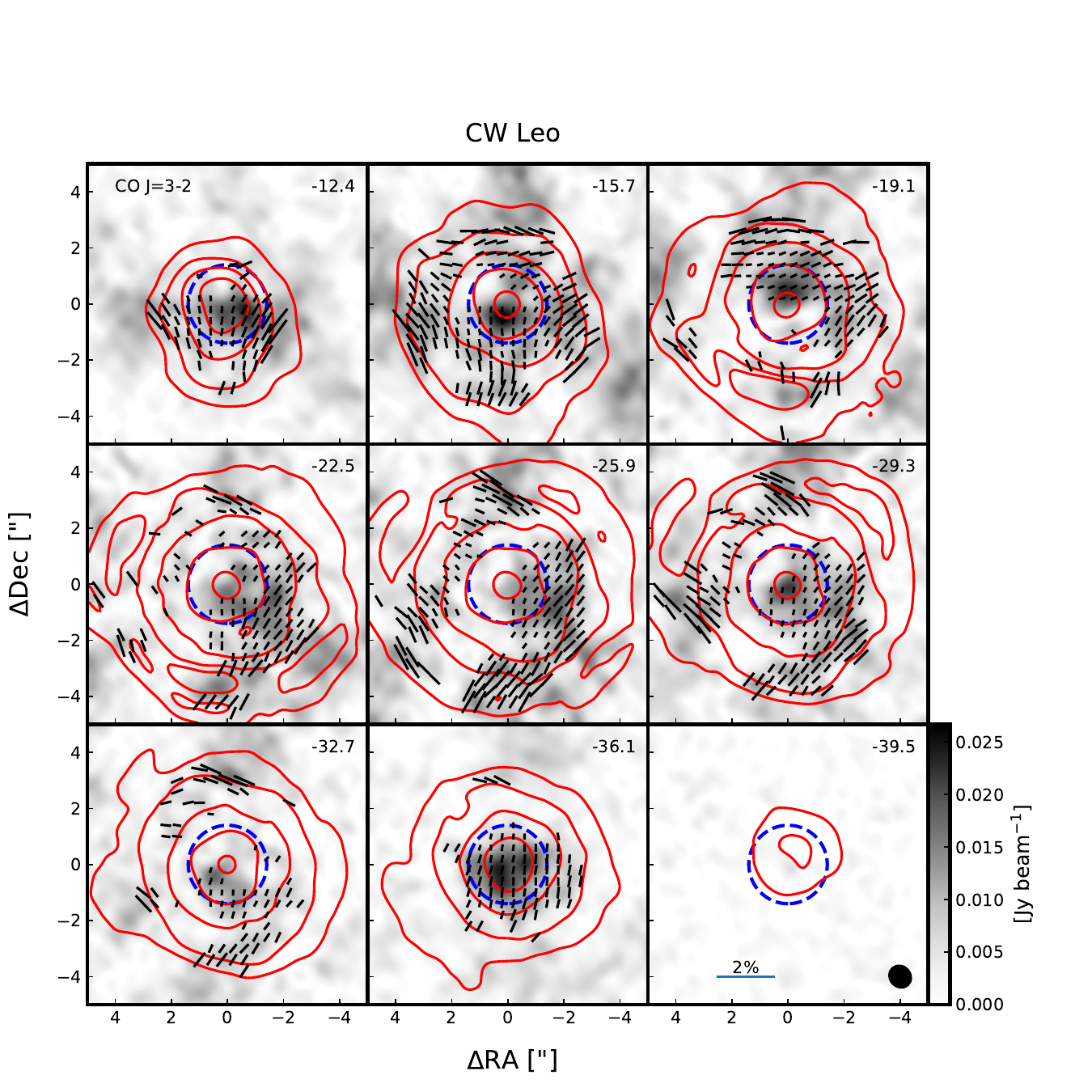}
\caption{Channel maps of the polarised CO $J=3-2$ emission around the AGB star CW~Leo. The solid red contours indicate the Stokes I total intensity emission at $5, 10, 20, 40,$ and $80\%$ of the peak emission ($I_{\rm CO, peak}$=17.89~Jy~beam$^{-1}$). The greyscale image is the linearly polarised emission, and the line segments denote the linear polarisation direction where emission is detected at $>5\sigma_P$. The segments are scaled to the level of fractional polarisation with the scale indicated in the bottom right panel. The maximum polarisation $P_{\rm l, max}= 1.18\%$. The beam size is denoted in the bottom right panel and all panels are labelled with the $V_{\rm lsr}$ velocity in \kms. The stellar velocity $V_{\rm lsr, *}=-26.5$~\kms. The dashed blue circles centred on peak of the CO emission close to the stellar velocity indicates the radius ($R\approx1.4$\arcsec) at which we find that the direction of polarisation for in particular CS changes from neither tangential nor radial to predominantly tangential.}
\label{IRCCO}
\end{figure*}

CW\,Leo is a C-type star located at a distance of 123\,pc \citep{Groenewegen12}. 
As the brightest source in the sky at 5\,$\mu$m, out of the Solar system, its circumstellar envelope has been source of numerous comprehensive studies, including the first detection of at least a fifth of the molecules known to exist in space \citep[see a recent detailed list in][]{Mcguire18}. 
This CSE has a ringed structure as seen in dust and molecular lines \citep{Mauron00,Leao06,Guelin18}. 
It has been formed as a consequence of a copious mass loss at an average rate of 2\,$\times$\,10$^{-5}$\,\my\ with enhanced episodic mass loss events, probably triggered by the presence of a binary companion \citep{Guelin18,Velilla19}. 
The CO envelope, as seen in CO $J$\,=\,2--1, extends up to 3\arcmin, equivalent to $\sim$\,3.3\,$\times$\,10$^{17}$\,cm at the adopted distance \citep{Guelin18}.
R\,Leo is a M-type star surrounded by a CSE of dust and, mainly, molecular gas formed at an average mass loss rate of 1\,$\times$\,10$^{-7}$\,\my\ \citep{2014A&A...566A.145R}.
Its distance is uncertain and it has been estimated to be in the range between approximately 70 to 110\,pc. 
The lower limit of 70\,pc comes from Gaia Data Release\,2 (DR\,2), which estimated a parallax of 14.06\,$\pm$\,0.84\,mas \citep{GaiaDR2}. 
The upper limit of 110\,pc was estimated by \cite{1995MNRAS.276..640H} applying the period-luminosity relation to their interferometric observational results with the William Herschel Telescope. 
Gaia estimates are considerably uncertain in the case of AGB stars \citep{Andriant22}.
In particular for R\,Leo, the fitting parameters of Gaia DR\,2 indicate a low quality fit and its solution should be then used with caution.
Moreover, despite new observations, no solution was found for R\,Leo in the recent Gaia DR\,3 \citep{GaiaDR3}. However, based on the Gaia observations of a large sample of AGB stars, \citet{Andriant22} have derived an updated period-luminosity relation which yields a distance to R~Leo of $100\pm5$~pc.
This is the distance we adopt for this source.
Based on the modelling of observations of CO $J$\,=\,6--5 with the Caltech Submillimeter Observatory and observations of lower excitation lines of CO from the literature, \cite{2006A&A...450..167T} estimated a CO photo-dissociation radius of 1.3\,$\times$\,10$^{16}$\,cm.

We selected these two sources, located sufficiently close on the sky to be observed with ALMA in a single observation, as representative cases of the C-type and M-type families to start the investigation of the effect of the chemistry (C-rich versus O-rich) on polarisation properties. However, for this, a study of a larger sample is required to arrive at any conclusions.

In \S~\ref{obs}, we present the ALMA observations and data reduction for both AGB stars. In \S~\ref{res} we present the observational results of the polarisation of $^{12}$CO, CS, H$^{13}$CN and SiS around CW~Leo and $^{12}$CO, $^{29}$SiO, H$^{13}$CN and SiO around R~Leo. The PORTAL models for both sources
are presented in \S~\ref{sec:polmods}. A discussion of the results, including a structure-function analysis to determine the magnetic field strength around CW~Leo, are given in \S~\ref{disc}. Finally, the conclusions are presented in \S~\ref{conc}.

\begin{figure*}[ht!]
\centering
\includegraphics[width=0.95\textwidth]{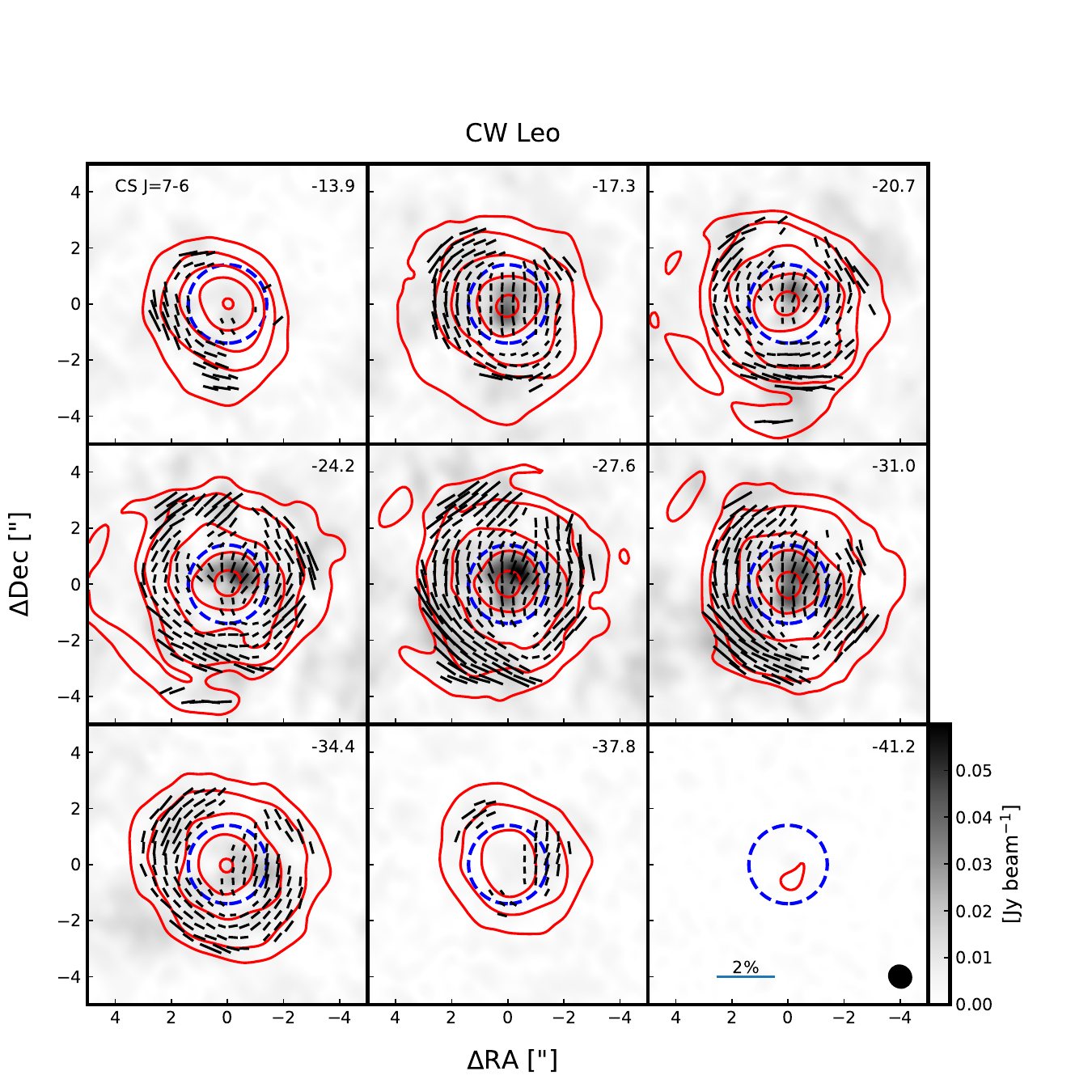}
\caption{Same as Fig.~\ref{IRCCO} for the CS $J=7-6$ emission around CW~Leo. The peak emission is $I_{\rm CS, peak}$=18.77~Jy~beam$^{-1}$. The maximum polarisation $P_{\rm l, max}= 1.74\%$.}
\label{IRCCS}
\end{figure*}

\begin{figure*}[ht!]
\centering
\includegraphics[width=0.95\textwidth]{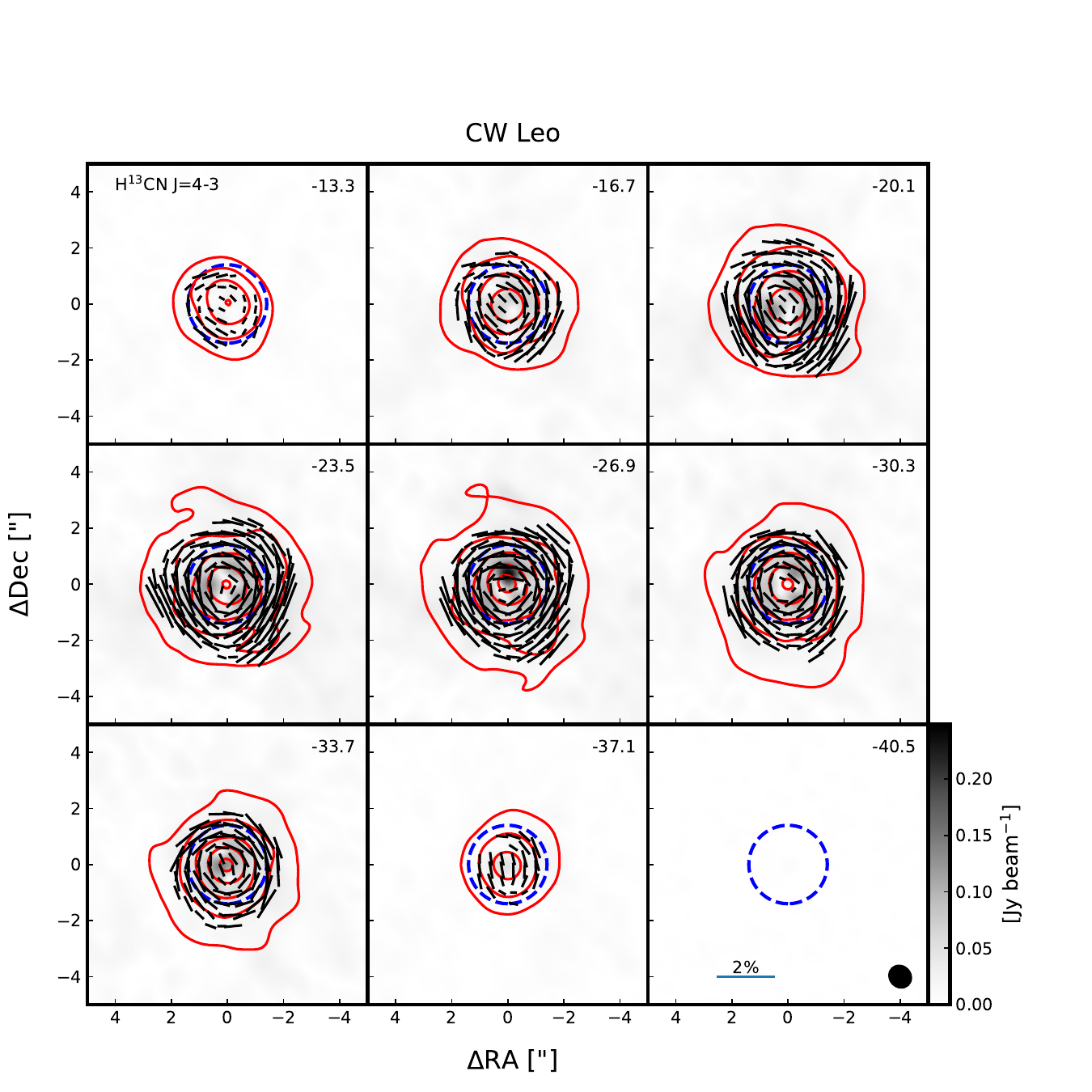}
\caption{Same as Fig.~\ref{IRCCO} for the H$^{13}$CN $J=4-3$ emission around CW~Leo. The peak emission is $I_{\rm HCN, peak}$=38.52~Jy~beam$^{-1}$. The maximum polarisation $P_{\rm l, max}= 1.62\%$.}
\label{IRCH13CN}
\end{figure*}

\begin{figure*}[ht!]
\centering
\includegraphics[width=0.95\textwidth]{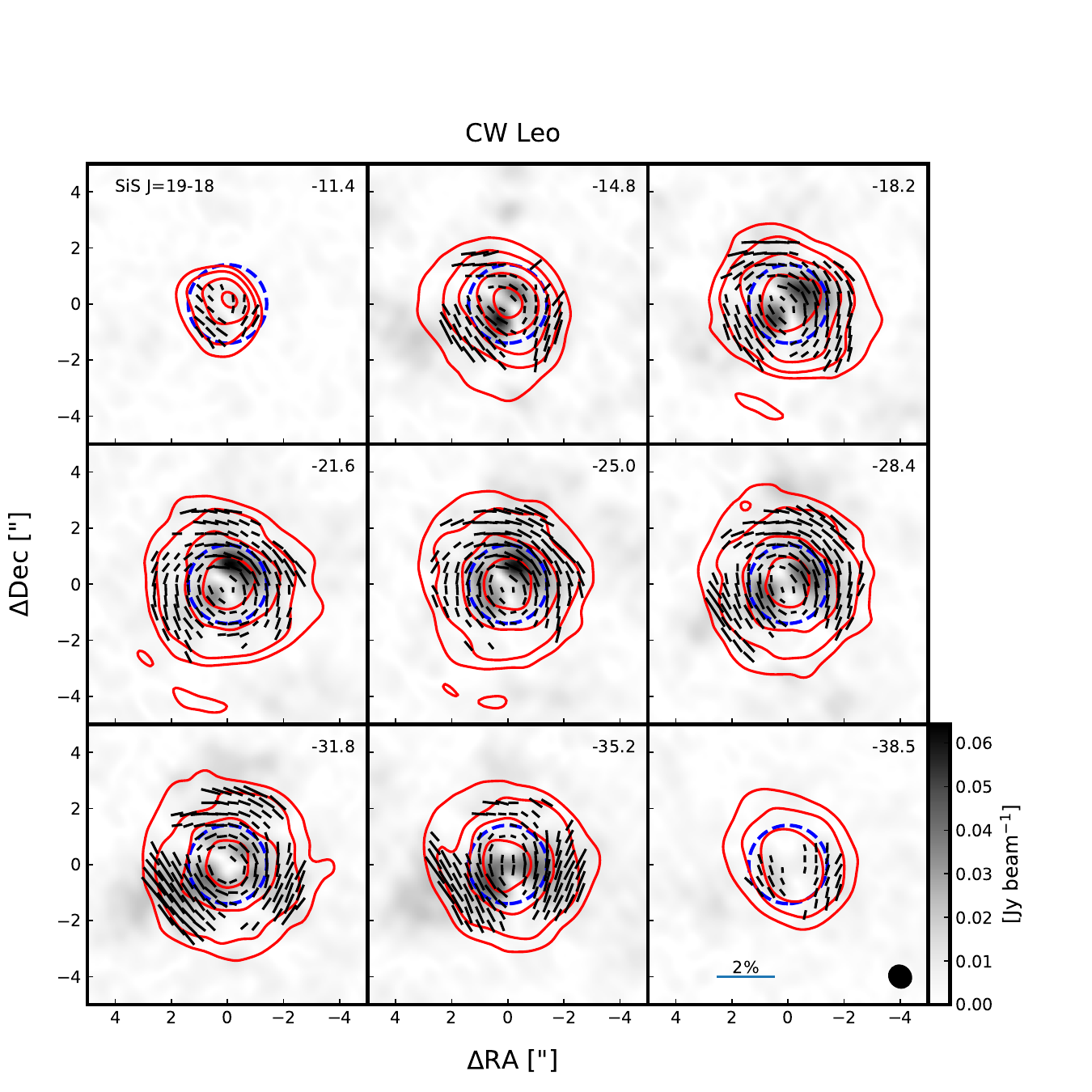}
\caption{Same as Fig.~\ref{IRCCO} for the SiS $J=19-18$ emission around CW~Leo. The peak emission is $I_{\rm SiS, peak}$=21.01~Jy~beam$^{-1}$. The maximum polarisation $P_{\rm l, max}= 1.47\%$.}
\label{IRCSiS}
\end{figure*}

\section{Observations and data reduction}
\label{obs}
The observations of CW~Leo and R~Leo were performed with ALMA in full polarisation mode on May 20 2018 (Project 2016.1.00251.S, PI:Vlemmings). Since both sources are relatively close to each other in the sky, both sources were observed using the same calibrators, spectral setup and in the same observing session of 3.3~hours. The total on source observing time for CW~Leo was $\sim 20$~min, while the time spent on R~Leo was $\sim 60$~min. The remaining time was used for observing the phase calibrator J1002+1216 and the amplitude and polarisation calibrator J0854+2006. The observations were done using 4 spectral windows (spw) of $1.875$~GHz with 960 spectral channels each. The spws were centred on $331.1$, $333.0$, $343.1$, and $345.0$~GHz and the resulting channel width was $\sim1.7$~\kms. Calibration was done using the ALMA polarisation calibration scripts \citep{Nagai16}. After the observations, an error in the visibility amplitude calibration was recognised in ALMA data of sources that contain strong line emission\footnote{\url{https://almascience.eso.org/news/amplitude-calibration-issue-affecting-some-alma-data}}. Considering CW~Leo fits this criterion, ALMA staff from the ESO ALMA Regional Centre performed a re-normalisation correction on our data. After that, calibration was redone using the standard procedure. A comparison between the results before and after re-normalisation indeed revealed significant differences in the total intensity spectra. The effect on the polarisation results was less significant and mostly affected the derived fractional linear polarisation. All results presented in this paper are based on the re-normalised data.

After the standard calibration, self-calibration and imaging was performed using CASA 5.7.2. Two rounds of phase-only self-calibration, (with solution intervals "inf" and "int") were done on the continuum. This increased the dynamic range on the CW~Leo continuum by a factor of $\sim5$. On the (weaker) continuum of R~Leo, the dynamic range improved by a factor of $\sim3.6$. The final continuum rms in Stokes I ($\sigma_I$) is $390~\mu$Jy~beam$^{-1}$ for CW~Leo and $98~\mu$Jy~beam$^{-1}$ for R~Leo. This corresponds to $\sim3-5$ times the theoretical noise limit and signal-to-noise ratios SNR$\sim 2720$ and $\sim3080$ for CW~Leo and R~Leo respectively. In the Stokes Q and U continuum images, the rms noise ($\sigma_{Q,U}$) for CW~Leo (R~Leo) are $\sim 100~(37)~\mu$Jy~beam$^{-1}$ and $\sim 130~(42)~\mu$Jy~beam$^{-1}$ respectively. The Stokes V rms noise ($\sigma_V$) in the continuum is  $\sim 90$ and $\sim 34~\mu$Jy~beam$^{-1}$ respectively for CW~Leo and R~Leo. The continuum beam sizes, using Briggs weighing and a robust parameter of $0.5$, are $0.79\times0.72$\arcsec (PA $39.0^\circ$) and $0.78\times0.71$\arcsec (PA $46.2^\circ$) for CW~Leo and R~Leo respectively. 

Before producing the spectral line cubes, the continuum was subtracted using the CASA task {\it uvcontsub}. Subsequently, the spectral lines were imaged using Briggs weighing and a robust parameter of $0.5$. For CW~Leo, we averaged two channels, obtaining a velocity resolution of $\sim 3.4$~\kms. No averaging was done for R~Leo, which was imaged at the native $1.7$~\kms~resolution. The $\sigma_I$, $\sigma_Q$, $\sigma_U$ and $\sigma_V$ rms noise level in a line free channel for CW~Leo (R~Leo) are $2.0~(1.1)~$mJy~beam$^{-1}$, $1.5~(1.1)~$mJy~beam$^{-1}$, $1.8~(1.2)~$mJy~beam$^{-1}$ and $1.6~(1.2)~$mJy~beam$^{-1}$. At 345~GHz, the beam sizes are $0.84\times0.75$\arcsec (PA $42.7^\circ$) and $0.83\times0.75$\arcsec (PA $49.1^\circ$) for CW~Leo and R~Leo respectively. The maximum recoverable scale in our observations is $\sim 7.9$\arcsec. For the most extended lines, such as $^{12}$CO $J=3-2$, this means that we resolve out a significant amount of flux. For example, compared to ALMA ACA observations of $^{12}$CO $J=3-2$ around R~Leo \citep{Ramstedt20}, we recover approximately half the Stokes I flux. For CW~Leo, a comparison with JCMT calibration observations indicate we resolve out $\sim70\%$ of the CO emission. As the polarised emission is often more compact, this can lead to an overestimate of the polarisation fractions with a factor of $2-3$, but this factor will vary over the image and depends on the exact morphology of both the total intensity and polarised intensity emission. Most of the other lines are much less affected. Compared to the ACA observations of R~Leo, the peak fluxes of H$^{13}$CN $J=4-3$ and $^{29}$SiO $J=8-7$ are $14\%$ and $7\%$ less respectively. Considering the absolute flux uncertainty in ALMA band 7 is $\sim10\%$, our observations thus likely recover most of the flux in these lines.   

Finally we produce debiased linear polarisation maps from the Stokes Q and U imaged using $P_l=\sqrt{Q^2+U^2-\sigma^2_P}$. The polarisation rms $\sigma_P$ is taken to be the mean of $\sigma_Q$ and $\sigma_U$. We also inspected the Stokes V circular polarisation image cubes and find that these are, for the extended molecular lines, dominated by ALMA beam effects \citep{Hull20}. For one of the more compact maser line, the V results appear reliable, but the limited spectral resolution does not allow for a detailed analysis of the Zeeman effect. Hence, when referring to polarisation in the rest of the paper, we mean linear polarisation unless otherwise mentioned.

\begin{figure*}
     \centering
     \begin{subfigure}[b]{0.48\textwidth}
         \centering
         \includegraphics[width=\textwidth]{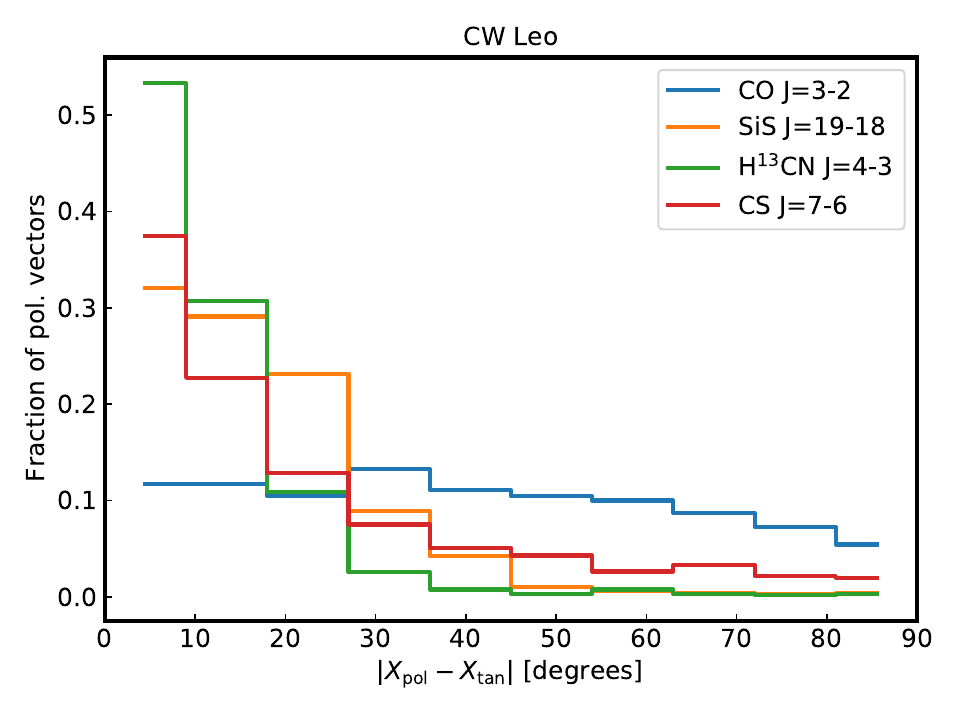}
     \end{subfigure}
     \hfill
     \begin{subfigure}[b]{0.48\textwidth}
         \centering
         \includegraphics[width=\textwidth]{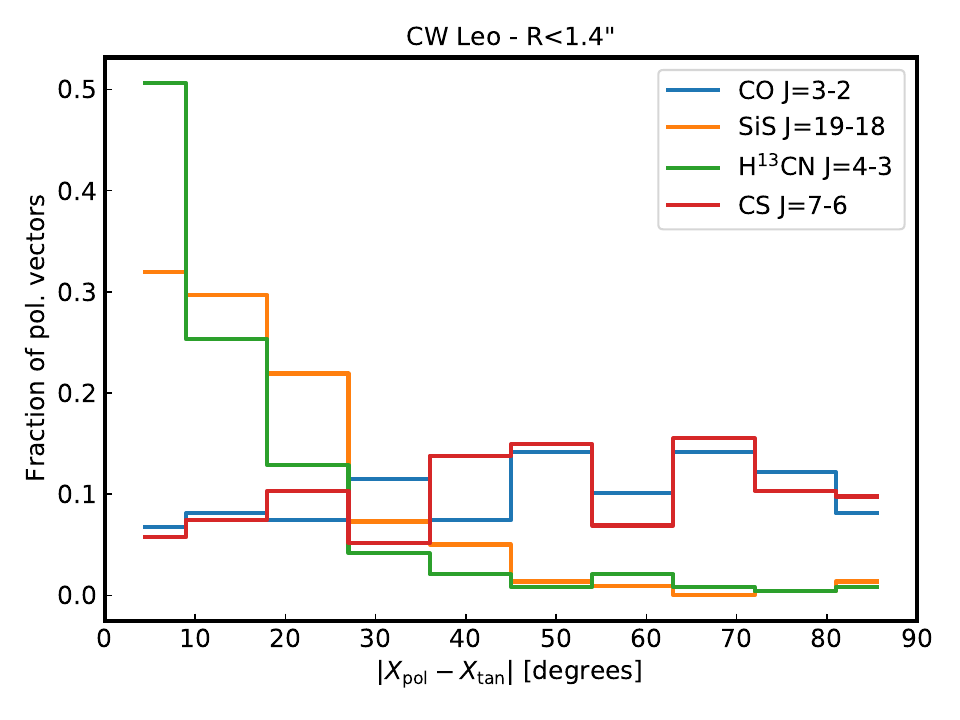}
     \end{subfigure}
     \hfill
        \caption{Distribution of polarisation angles with respect to the tangential direction for the emission of CO $J=3-2$, SiS $J=19-18$, H$^{13}$CN $J=4-3$, and CS $J=7-6$ around CW~Leo. {\it (left)} The polarisation angles taken from the entire circumstellar envelope emission. {\it (right)} The polarisation angles for $R<1.4$\arcsec. }
        \label{fig:angdist}
\end{figure*}

\begin{figure*}[ht!]
\centering
\includegraphics[width=0.95\textwidth]{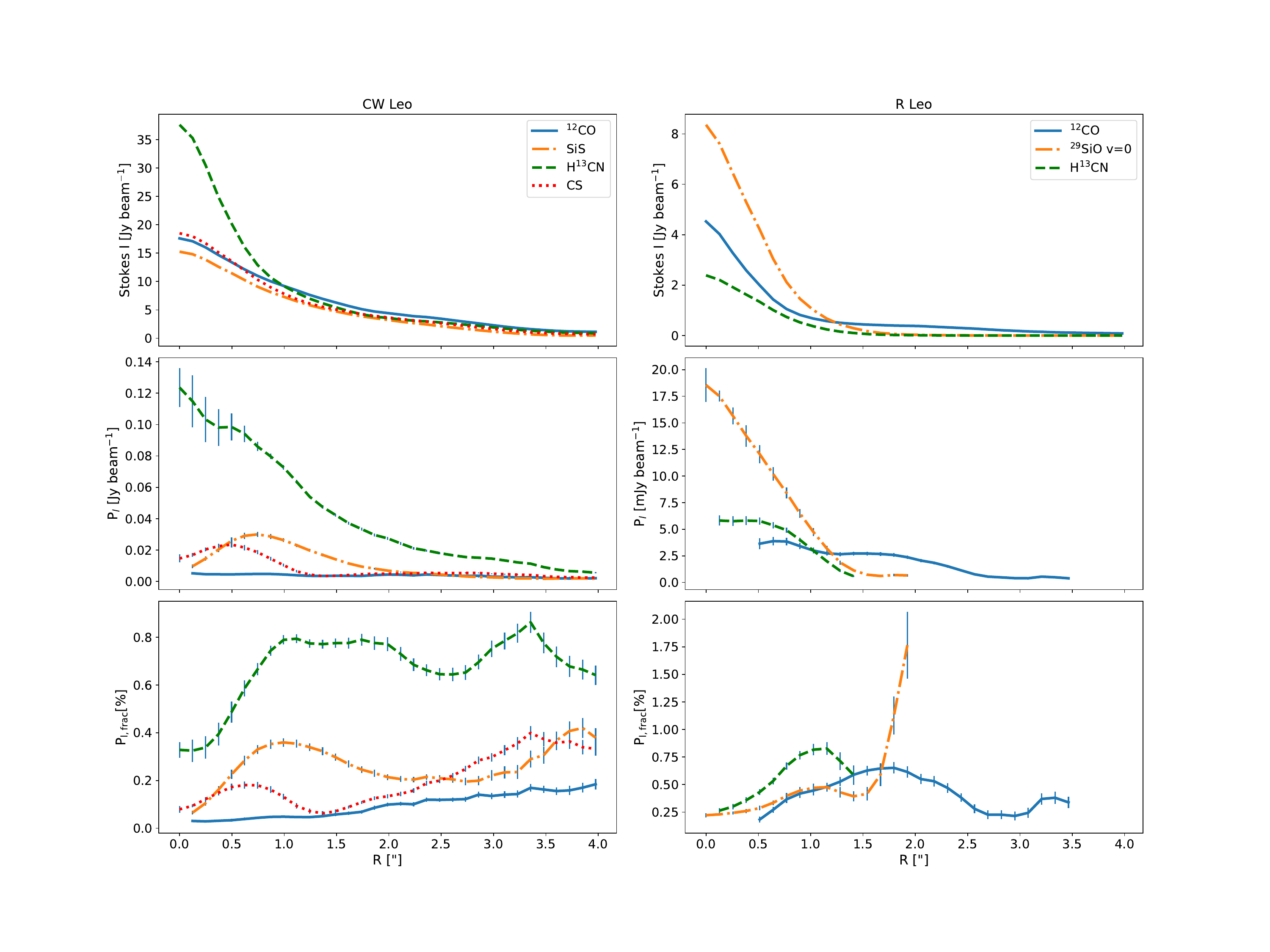}
\caption{Azimuthally averaged Stokes I (top), polarised intensity $(P_l)$ (middle) and polarisation fraction $(P_{\rm l,frac})$ (bottom), taken in the channel that includes the stellar velocity, for CW~Leo (left column) and R~Leo (right column). For CW~Leo, we show CO $J=3-2$, SiS $J=19-18$, H$^{13}$CN $J=4-3$, and CS $J=7-6$. For R~Leo we show CO $J=3-2$, $^{29}$SiO $v=0, J=4-3$, and H$^{13}$CN $J=4-3$. We did not include the mostly unresolved SiO $v=1,$ and $v=2, J=8-7$ transitions. The profiles of $P_l$ and $P_{\rm l,frac}$ include the error bars from the azimuthal averaging and only radial points for which the  SNR$>5$ are plotted. Because of the averaging, the polarisation fraction is much less than the maximum fraction detected in the image cubes for each line. Beyond $1/3$rd the ALMA primary beam size (at $R\approx3.0$) systematic errors might start to contribute to the polarisation signal. Note that the units for the polarised intensity is different for the two middle plots for CW~Leo (Jy~beam$^{-1}$) and R~Leo (mJy~beam$^{-1}$).}
\label{Fig:Radial}
\end{figure*}

\begin{figure*}[ht!]
\centering
\includegraphics[width=0.95\textwidth]{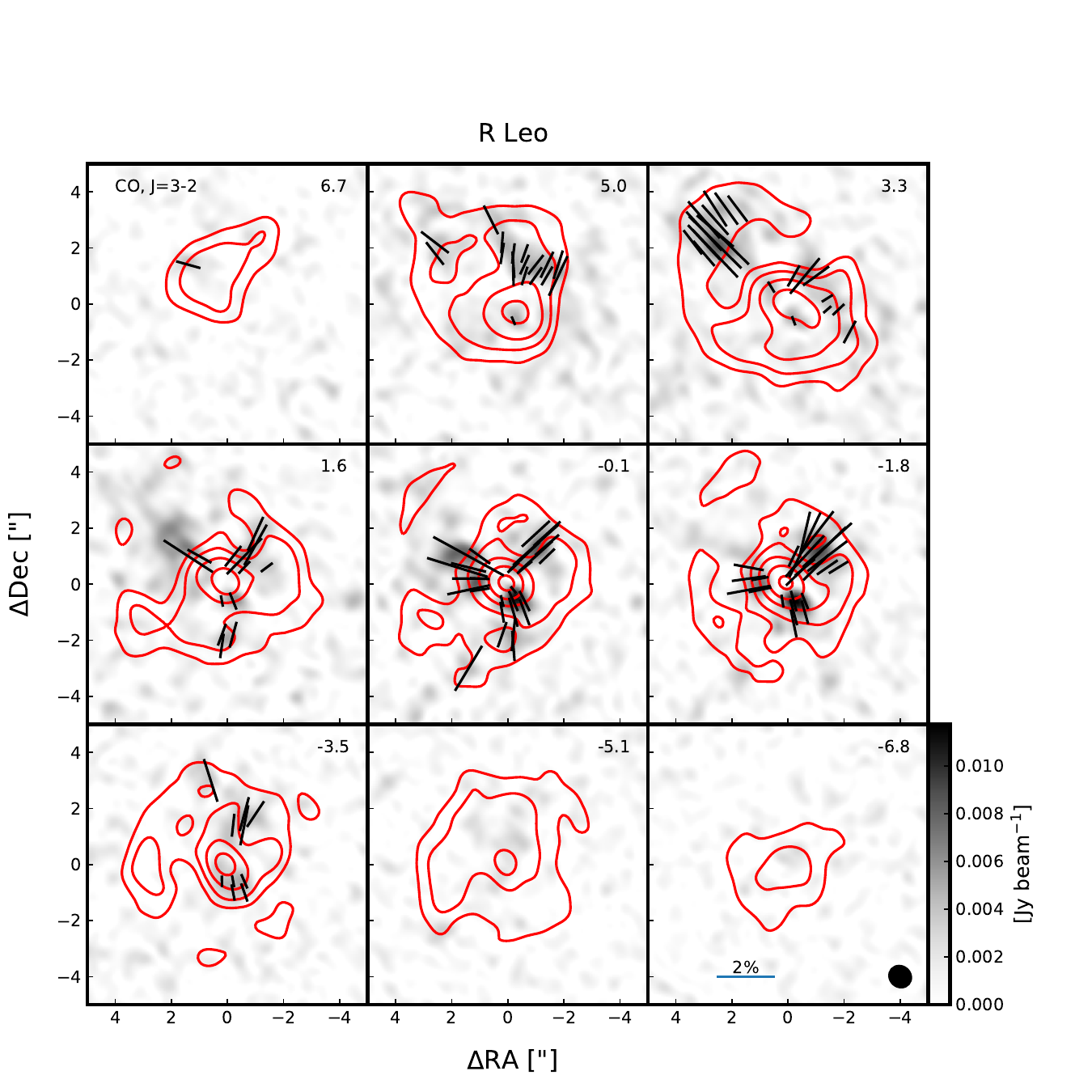}
\caption{Same as Fig.~\ref{IRCCO} for the CO $J=3-2$ emission around R~Leo. The stellar velocity $V_{\rm lsr, *}=-0.5$~\kms. The peak emission is $I_{\rm CO, peak}$=4.66~Jy~beam$^{-1}$. The maximum polarisation $P_{\rm l, max}= 3.90\%$.}
\label{RLeoCO}
\end{figure*}

\section{Observational results}
\label{res}

\subsection{Polarisation result overview}
We detect significant ($>5\sigma_P$) polarised emission towards the 338~GHz continuum of both CW~Leo and R~Leo, which we show in Fig.~\ref{fig:contpol}. The maximum fraction of continuum polarisation is $0.88\%$ for CW~Leo and $0.17\%$ for R~Leo. We detect no significant circular polarisation signal in the Stokes V continuum image, and can place a $5\sigma_V$ limit on the continuum circular polarisation of $0.05\%$ for CW~Leo and $0.02\%$ for R~Leo. The continuum emission of both stars is dominated by the free-free emission from the extended stellar atmosphere. For R~Leo, comparing with the high angular resolution ALMA observations at $\sim230$~GHz from \citep{Vlemmings19}, and using a spectral index of $1.8$ consistent with the measurements, yields an expected stellar continuum flux of $\sim225$~mJy. Considering we measure an integrated continuum flux of $318$~mJy, dust emission contributes approximately $30\%$ to the continuum at $338~$GHz. If the polarisation of the continuum of R~Leo is due to dust polarisation, this means the dust fractional polarisation is $\sim0.6\%$.
Recent ALMA high angular resolution observations of CW~Leo at 258~GHz found a stellar continuum flux of $\sim500$~mJy \citep{Velilla23}.
Extrapolating this measurement with a spectral index between of $1.8$, taken the same as R~Leo, this means the stellar flux of CW~Leo at 338~GHz is $\sim810$~mJy.
Considering the integrated continuum flux of $1.143$~Jy, this would also imply a dust contribution of $30\%$. Thus, if the continuum polarisation is only due to dust polarisation, the dust polarisation fraction is $3\%$. Because of the large uncertainty, the lack of spatial resolution as well as the lack of multi-frequency observations, we can not determine the processes responsible for the dust polarisation. Hence we did not perform any further analysis on the continuum polarisation.

In addition to the continuum polarisation, we measured polarisation at a level of more than $5\sigma_P$ for four molecular transitions around CW~Leo and for five molecular transitions around R~Leo. The peak fluxes ($I_{\rm peak}$) of the lines for which polarisation was detected, as well as the maximum polarisation fraction found in the molecular line maps ($P_{\rm l, max}$) are presented in Table~\ref{Table:res}. As discussed before, the fractional polarisation of the extended emission lines can be affected by filtering out the large scale structure. We now discuss the molecular line polarisation of both sources in more detail.

\begin{table}
\caption{Polarisation observational results}             
\label{Table:res}      
\centering          
\begin{tabular}{l c  c }     
\hline\hline
Molecular Line & $I_{\rm peak}$ & $P_{\rm l, max}$ \\
 & [Jy~beam$^{-1}$] & [$\%$] \\
 \hline
\multicolumn{3}{c}{CW~Leo}\\
\hline
$^{12}$CO $J=3-2$ & 17.89 & 1.18 \\
CS $J=7-6$ & 18.77 & 1.74 \\
H$^{13}$CN $J=4-3$ & 38.52 & 1.62 \\
SiS $J=19-18$ & 21.06 & 1.47 \\
\hline
\multicolumn{3}{c}{R~Leo}\\
\hline
$^{12}$CO $J=3-2$ & 4.66 & 3.90 \\
$^{29}$SiO $v=0,J=8-7$ & 8.49 & 1.57 \\
H$^{13}$CN $J=4-3$ & 2.40 & 1.79 \\
SiO $v=1,J=8-7$ & 12.68 & 32.4 \\
SiO $v=2,J=8-7$ & 0.72 & 4.92 \\
\hline\hline       
\end{tabular}
\end{table}

\subsection{CW~Leo}
\label{obs: CW Leo}

We detect significant polarisation around CW~Leo for four molecular lines. The emission maps of $^{12}$CO $J=3-2$, CS $J=7-6$, H$^{13}$CN $J=4-3$, and SiS $J=19-18$ are shown in Figs.~\ref{IRCCO},~\ref{IRCCS},~\ref{IRCH13CN}, and~\ref{IRCSiS} respectively. The maximum fractional polarisation for the four lines ranges are all between $1-2\%$ but as can be seen in the images, the distribution of the polarised emission and the orientation of the polarisation vectors is different for the different lines. While the polarisation vectors of the H$^{13}$CN $J=4-3$ and SiS $J=19-18$ lines are mainly in the tangential direction, those in the CS $J=7-6$ line display a direction that is neither tangential nor radial for $R<1.4$\arcsec. We thus interpret $R=1.4$\arcsec as the radius where in particular the alignment mechanism for CS undergoes a change. The $^{12}$CO $J=3-2$ shows a polarisation structure that is different from all three other lines at larger ($R>1.4$\arcsec) distances from the centre of the emission and a structure that is consistent with that of CS $J=7-6$ in the inner envelope. To illustrate the different behaviour of the polarisation direction we plot a histogram with the deviation from the tangential direction for all polarisation vectors in Fig.~\ref{fig:angdist}(left) and limited to those vectors within $R<1.4$\arcsec~in Fig.~\ref{fig:angdist}(right). 
All lines show a good correspondence of polarisation structure between neighbouring channels, despite that the possible effect of the resolved-out large scale emission is different for the different velocity channels and should be most pronounced closer to the stellar velocity. It is thus unlikely that the observed polarisation structure itself is strongly affected by missing flux. 

In Fig.~\ref{Fig:Radial}(left) we also present the azimuthally averaged Stokes I, linearly polarised flux and fractional polarisation for the spectral channel closest to the stellar velocity ($V_{\rm lsr, CW Leo}=-26.5$~\kms). Here we note that, for CW~Leo, the relative polarisation of the H$^{13}$CN line is stronger than that of the other lines. As expected because of the averaging of polarisation structure within the observing beam, all lines show a decrease of fractional polarisation in the central beam. Outside of this, the polarisation fraction of H$^{13}$CN remains fairly stable around $0.7\%$ while the average polarisation fraction of $^{12}$CO is much lower, but rises steadily outwards. The fractional polarisation of CS shows a minimum around $\sim 1.4$\arcsec, which is consistent with the observed change in morphology at that radius. Finally, the fractional polarisation of SiS behaves somewhat similar to that of H$^{13}$CN although it shows a somewhat more pronounced peak around $\sim 1$\arcsec before decreasing to $\sim 0.2\%$ and finally, beyond $\sim 3$\arcsec rising again to peak at $\sim 0.4\%$.

\begin{figure*}[ht!]
\centering
\includegraphics[width=0.95\textwidth]{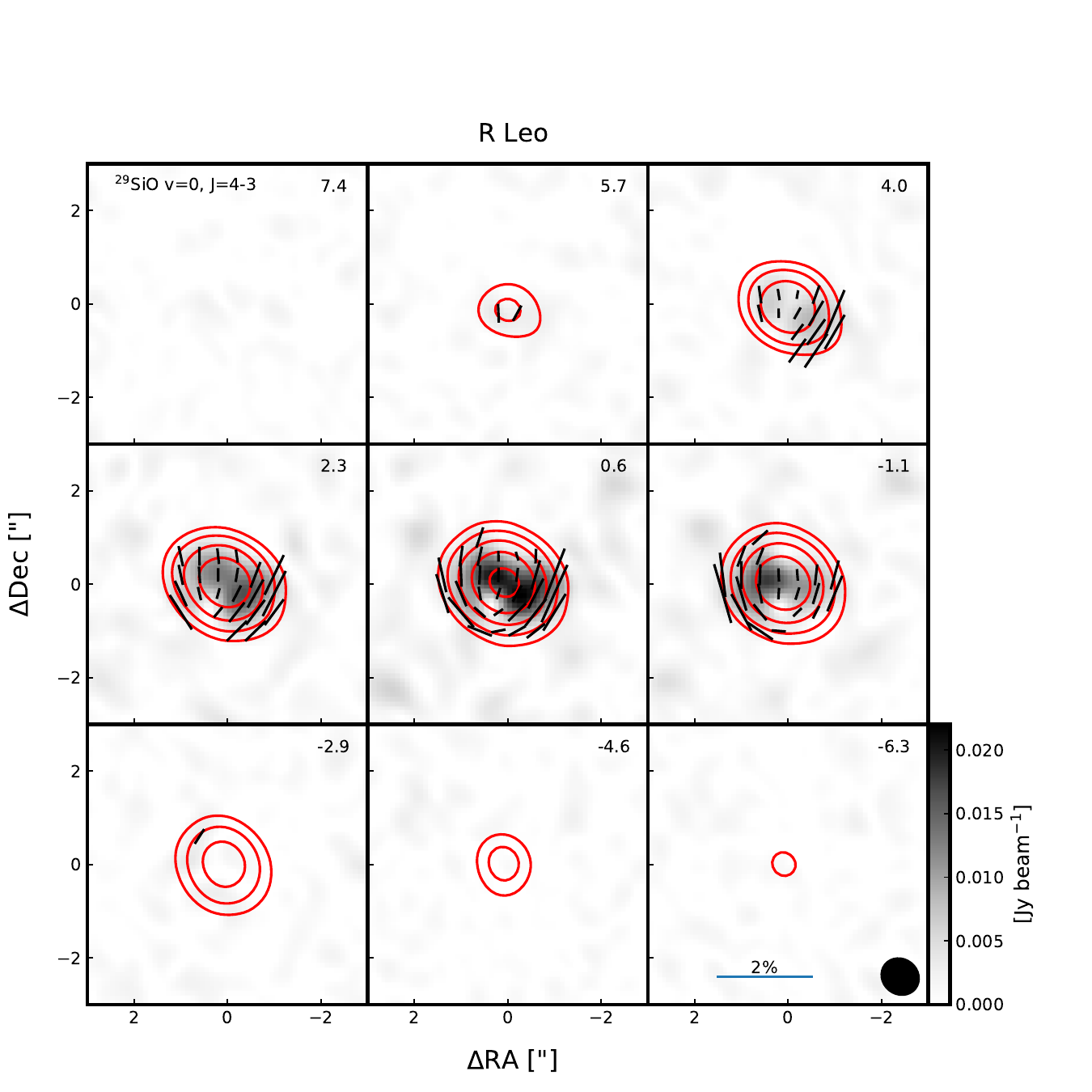}
\caption{Same as Fig.~\ref{IRCCO} for the $^{29}$SiO $v=0, J=8-7$ emission around R~Leo. The peak emission is $I_{\rm 29SiO, peak}$=8.49~Jy~beam$^{-1}$. The maximum polarisation $P_{\rm l, max}= 1.57\%$.}
\label{RLeo29SiO}
\end{figure*}

\begin{figure*}[ht!]
\centering
\includegraphics[width=0.95\textwidth]{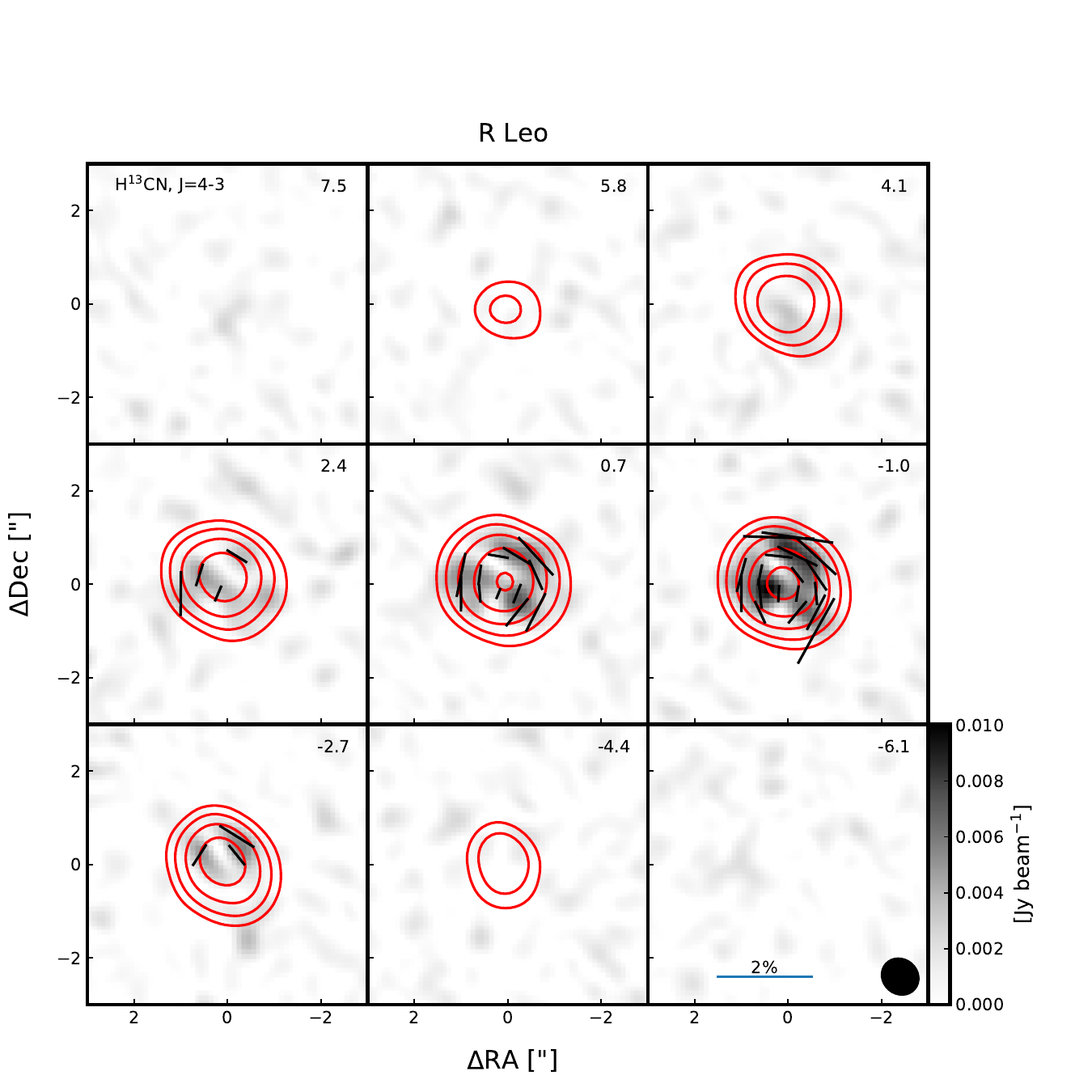}
\caption{Same as Fig.~\ref{IRCCO} for the H$^{13}$CN $J=4-3$ emission around R~Leo. The peak emission is $I_{\rm HCN, peak}$=2.40~Jy~beam$^{-1}$. The maximum polarisation $P_{\rm l, max}= 1.79\%$.}
\label{RLeoH13CN}
\end{figure*}

\subsection{R~Leo}

 Around R~Leo we find polarisation for $^{12}$CO $J=3-2$, $^{29}$SiO $v=0, J=8-7$, and H$^{13}$CN $J=4-3$, as well as for the $v=1$ and $v=2$ SiO $J=8-7$ transitions. Maps for the first three of these are presented in Figs.~\ref{RLeoCO},~\ref{RLeo29SiO}, and~\ref{RLeoH13CN} respectively. Maps for the vibrationally excited transitions, that likely include maser amplification, are presented in Figs.~\ref{RLeoSiOv1} and \ref{RLeoSiOv2}. While the peak level of fractional polarisation of the H$^{13}$CN around R~Leo is similar to that around CW~Leo, the peak fractional polarisation of the $^{12}$CO is three times higher. As the level of resolved out flux is larger for the extended envelope of CW~Leo, which means that its polarisation fraction is likely more overestimated than for R~Leo, the difference between the peak fractional polarisation of the $^{12}$CO of both sources is intrinsic. The peak fractional polarisation of the SiO vibrationally excited lines are higher still, with that of the $v=1$ transition reaching a level above $30\%$. This is consistent with previous observations of high frequency SiO masers \citep[e.g.][]{Vlemmings11, Vlemmings17a} and with theoretical predictions \citep[e.g.][]{Lankhaar19, Lankhaar24}.
 
\begin{figure}[ht]
         \centering
         \includegraphics[width=0.45\textwidth]{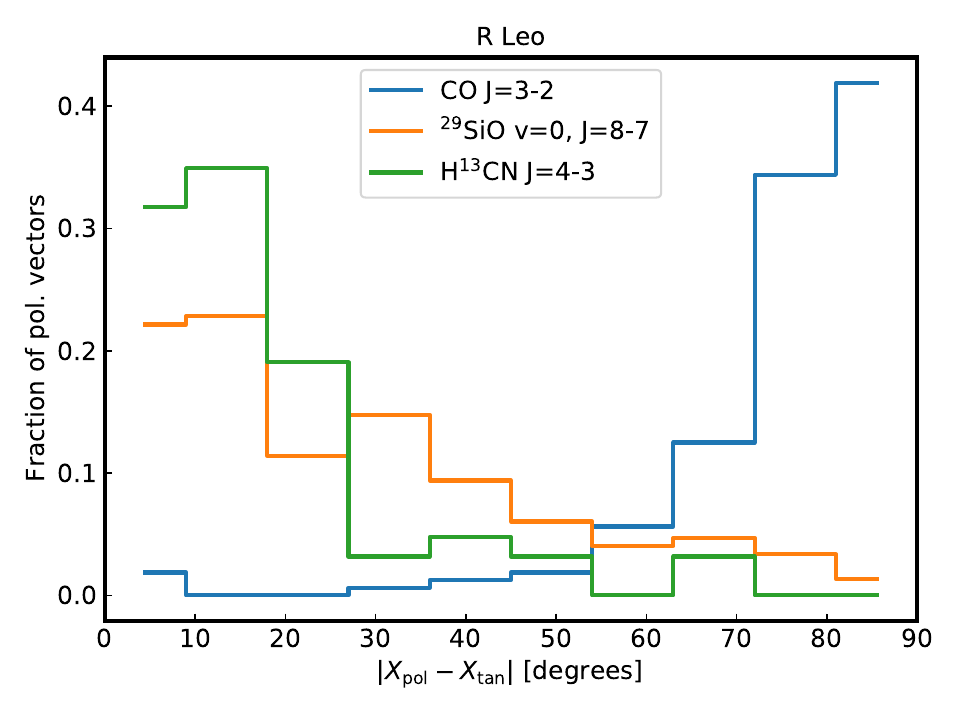}
        \caption{Distribution of polarisation angles with respect to the tangential direction for the emission of CO $J=3-2$, $^{29}$SiO $v=0, J=8-7$, and H$^{13}$CN $J=4-3$ around R~Leo.}
        \label{fig:angdist_RLeo}
\end{figure}

 At the same time, the direction of the polarisation vectors of $^{12}$CO around R~Leo are clearly radial. As illustrated in Fig.~\ref{fig:angdist_RLeo}, the vectors of $^{29}$SiO and H$^{13}$CN are mostly tangential. For the likely masing vibrationally excited states of SiO, the detected polarisation is concentrated towards the central emission peak. This is consistent with the polarised emission originating from masers close to the central star. From positive to negative $V_{\rm lsr}$, the polarisation vectors position angle of the $v=1$ transition rotates smoothly from $36.4\pm0.9^\circ$ to $26.3\pm1.0^\circ$ while that of the $v=2$ transition rotates from $47\pm2^\circ$ to $22\pm3^\circ$. This indicates a preferred direction of the polarised emission of SiO close to the star, but our angular resolution is not sufficient to draw any stronger conclusions. As noted before, we do detect circularly polarised emission of the SiO $v=1$ maser line with a (negative) peak flux of $I_v=-164\pm1$~mJy~beam$^{-1}$. This corresponds to a circular polarisation percentage of $P_v=1.3\%$. As the intrinsic maser velocity width is much smaller than the spectral channel width, this percentage will be a lower limit to the actual circular polarisation. Assuming the relation between SiO maser circular polarisation fraction and magnetic field strength from \citet{1997ApJ...481L.111K}, this corresponds to a magnetic field of $B\sim2-3$~G in the SiO maser region, which is consistent with previous measurements \citep{Herpin06}. No circular polarisation above a $5\sigma_V$ level of $5.6$~mJy~beam$^{-1}$ is seen for the significantly weaker $v=2$ transition.   
 
 The azimuthally averaged Stokes I, linear polarisation and fractional polarisation profiles for R~Leo are shown in Fig.~\ref{Fig:Radial}(right). Also for the average polarisation, it is clear that the CO fractional polarisation of R~Leo is higher than that of CW~Leo, while the H$^{13}$CN polarisation levels are similar. All lines show the reduced fractional polarisation towards the centre and for both $^{29}$SiO and H$^{13}$CN the fractional polarisation appears to peak around $R\sim 1.1$\arcsec~ before slightly decreasing. After this decrease, the fractional polarisation of $^{29}$SiO rises sharply towards the outer edge of the $^{29}$SiO envelope while for H$^{13}$CN the polarised emission becomes too weak to be detectable. The CO polarisation extends further but the fractional polarisation reduces after peaking at $R\sim1.7$\arcsec.

\section{Polarised radiative transfer models}
\label{sec:polmods}
We simulated the emergence of polarisation of the emission from different molecules in the circumstellar envelope towards CW Leo and R Leo. The modelling of the line polarisation is performed with the PORTAL (POlarised Radiative Transfer Adapted to Lines) code \citep{Lankhaar20}. PORTAL uses the anisotropic intensity approximation \citep[for a thorough discussion, see][]{Lankhaar20} and assumes a dominant symmetry axis for the molecular species of interest. In the case of molecular lines excited in the CSE around evolved stars, the dominant symmetry axis may either be aligned with the magnetic field, or the infrared radiation field, which is typically radial. Radiative alignment is expected when radiative interactions occur at a higher rate than the magnetic precession rate ($\sim\mathrm{s}^{-1}/\mathrm{mG}$).

PORTAL maps out the intensity and anisotropy of the radiation field, at the resonant frequencies of the molecule under investigation, throughout the simulation. The radiative transfer is performed using the converged output of a LIME (version 1.9.5) simulation \citep{Brinch10}. The radiation field (anisotropy) parameters, in conjunction with the magnetic field geometry and density profile, are subsequently used to model the molecular excitation and quantum state alignment throughout the simulation. With the molecular excitation and alignment parameters, one is able to perform a polarised ray-tracing to produce a polarised image of the region of interest.

We modelled the emergence of line polarisation in CO, CS, H$^{13}$CN, SiS and $^{29}$SiO. These molecules are exposed to a strong radiation field from the central AGB star, and as a result, transitions involving the vibrationally excited states occur at high rates, thus having a high tendency to align the molecular population. We have therefore also included the first vibrationally excited states in modelling the polarised line emission of these molecules. For CO, we include the $J=0-40$ levels of the first two vibrational states, using the transition frequencies and Einstein coefficients from \citet{li:cdms} and the collisional rate coefficients from \citet{castro:17}, assuming an ortho-to-para ratio of 3:1. For CS, we augmented the standard LAMDA datafile \citep{lamda}, that include collisional rate coefficients of \citet{lique:06}, with the first vibrationally excited state, including only its radiative coupling to the ground state and taking the radiative rates from \citet{li:cdms}. We used the data from \citet{danilovich:13} to model H$^{13}$CN, where we limited the vibrational excitation to the first excited bending mode. For SiS, we augmented the standard LAMDA datafile \citep{lamda}, that include collisional rate coefficients of \citet{dayou:06}, with the first vibrationally excited state, including only its radiative coupling to the ground state and taking the radiative rates from \citep{li:cdms}. Finally, to model $^{29}$SiO, we used the molecular data file from \citet{danilovich:13}.

\subsection{Circumstellar envelope models}
We performed radiative transfer simulations of circumstellar envelopes of CW\,Leo and R\,Leo. Parameterised models of the physical conditions relevant to the radiative transfer towards these objects have been used. The gas density profile is modelled, assuming a spherically symmetric envelope of molecular gas and dust expanding at constant velocity \citep[see e.g.][and references therein]{Velilla19}
\begin{equation}
    n\,=\,\frac{\dot{M}}{4\pi\,r^2\,v_{\infty}\,\langle m_g \rangle},
    \label{Eq:n}
\end{equation}
where $\dot{M}$ is the mass-loss rate, $r$ is the radial distance to the star, $v_\infty$ is the terminal expansion velocity, and $\langle m_g \rangle$ is the average mass of particles in the gas, for which we adopted 2.35\,AMU based on a gas primarily made up of molecular hydrogen and helium. The gas kinetic temperature for CW Leo was assumed to follow the profile \citep{2012A&A...543A..48A} 
\begin{subequations}
\label{eq:tgas CW Leo}
\begin{align}
T_{\mathrm{k,gas}}  = \begin{cases}
T_{*}\,(r/R_* )^{-0.55} \quad &\mathrm{for}\ r \leq 75\ R_* \\
T(75\ R_* )\,(r/75\ R_* )^{-0.85}\quad &\mathrm{for}\ 75\ R_*  < r \leq 200\ R_*, \\
T(200\ R_*)\,(r/200\ R_* )^{-1.4} \quad &\mathrm{for}\ r > 200\ R_* ,
\end{cases}
\end{align}
where $T_{*}$ stands for the stellar temperature and $R_*$ is the stellar radius. For R Leo, we adopted the kinetic gas temperature profile,
\begin{align}
\label{eq:tgas R Leo}
T_{\mathrm{k,gas}}  = 
T_{*}\,(r/R_{*})^{-0.65},
\end{align}
\end{subequations}
which was based on the kinetic temperature profile for the similar relatively low mass-loss rate AGB star W Hya \citep{Khouri14}. For both CW Leo and R Leo, we assumed a lower limit on the gas kinetic temperature of $10$ K. The gas temperature was used to compute the velocity dispersion, that furthermore contained a contribution from a constant turbulent velocity of 1\,\kms. The dust temperature was assumed to follow
\begin{align}
\label{eq:tdust}
T_{\mathrm{k,dust}}  = \begin{cases}
T_{\mathrm{c}} \quad &\mathrm{for}\ r \leq R_{\mathrm{c}} \\
T_{\mathrm{c}}\,(r/R_{\mathrm{c}} )^{-q_{\mathrm{c}}} \quad &\mathrm{for}\ r > R_{\mathrm{c}}\end{cases},
\end{align}
where $R_{\mathrm{c}}$ is the dust condensation radius, $T_{\mathrm{c}}$ the dust condensation temperature and $q_{\mathrm{c}}$ is the parameterised power-law exponent describing the drop-off of the dust temperature. We did not find a dust temperature profile specifically tailored to R~Leo but instead adopted the the condensation temperature and radius from \citet{2014A&A...566A.145R} and the exponent from an average for M-type stars from \citet{1997A&A...322..924M}. In LIME, the dust density is determined from the gas density using the gas-to-dust ratio. To enforce the absence of dust within the dust condensation radius, for $r<R_{\mathrm{c}}$ the gas-to-dust mass ratio was set to the arbitrarily high value of $10^8$. 
The radiative transfer is highly dependent on the type of dust that is present in the CSE. In the case of C-rich CSEs, such as CW\,Leo, we adopted the opacities for amorphous carbon type of dust given by \cite{suh00}, while for O-rich CSEs, such as R\,Leo, we adopted the opacities for silicate type from \cite{suh99}. 

The gas velocity profile was assumed to only have a component in the radial direction. For the radial velocity profile, $v_{\mathrm{exp}}$, we assumed for CW Leo the profile,
\begin{subequations}
\label{eq:vexp}
\begin{align}
v_{\mathrm{exp}} = \begin{cases}
v_{\infty} \left(1-\left[\frac{R_{\mathrm{w}}-r}{R_{\mathrm{w}}}\right]^{2.5}\right) \quad &\mathrm{for}\ r \leq R_{\mathrm{w}} \\
v_{\infty} \quad &\mathrm{for}\ r > R_{\mathrm{w}} \end{cases},
\end{align}
where $R_{\mathrm{w}}$ is the wind acceleration radius and $v_{\infty}$ the terminal wind velocity. We based the CW Leo expansion profile on \citet{2012A&A...543A..48A}. For R Leo, we adopted a velocity profile similar to W Hya \citep{Khouri14},
\begin{align}
v_{\mathrm{exp}} =
3\ \mathrm{km \ s^{-1}} + (v_{\infty}-3\ \mathrm{km \ s^{-1}} ) \left(1-\frac{3.68\, \mathrm{AU}}{r}\right)^{5} .
\end{align}
\end{subequations}
The parameters that were used to represent CW Leo- and R Leo-like circumstellar envelopes are summarised in Table\,\ref{tab:agbmods}.
We have attempted to be as representative as possible to the models that exist in the literature but have not adjusted the model parameters for the different distances that were used to derive some of the parameters.

\begin{table*}[hbtp!]
\caption{Stellar envelope parameters}
\label{tab:agbmods}
\begin{center}
\begin{tabular}{l c c}
\hline\hline
          & CW Leo & R Leo             \\
\hline              
Distance ($d$)                                           & 123\,pc$^{(a)}$  & 100\,pc$^{(b)}$                   \\
Stellar radius ($R_\mathrm{*}$)                          & 2.7\,AU$^{(c)}$    & 1.4\,AU$^{(d)}$	               \\
Stellar effective temperature ($T_\mathrm{*}$)           & 2330\,K$^{(c)}$  & 2570\,K$^{(d)}$                   \\
Mass loss rate ($\dot{M}$)                               & $2\times 10^{-5}$\,\my$^{(c)}$   & $10^{-7}$\,\my$^{(e)}$	               \\
Terminal expansion velocity ($v_{\infty}$)                                 & 14.5\,\kms$^{(c)}$ &  6\,\kms$^{(e)}$                                \\
Dust condensation temperature ($T_\mathrm{c}$)                & 800\,K$^{(c)}$ & 1200\,K$^{(e)}$    \\
Dust condensation radius ($R_\mathrm{c}$)                & 13.5\,AU$^{(e)}$ & 8.7\,AU$^{(e)}$    \\
Dust temperature exponent ($q_\mathrm{c}$)                & 0.375$^{(c)}$ & 0.34$^{(f)}$    \\
Wind acceleration radius ($R_\mathrm{w}$)                & 20\,$R_\mathrm{*}$$^{(c)}$ & $-$    \\
Gas-to-dust mass ratio ($\rho_\mathrm{g}/\rho_\mathrm{d}$)(r\,$>$\,$R_\mathrm{c}$) & 300$^{(c)}$ & 167$^{(g)}$               \\      
\hline                                                      
\hline
\end{tabular}
\end{center}
\tablebib{
$^{(a)}$\citet{Groenewegen12}; $^{(b)}$\citet{Andriant22}; $^{(c)}$\citet{2012A&A...543A..48A};  $^{(d)}$\citet{Wittkowski16}; $^{(e)}$\citet{2014A&A...566A.145R}; $^{(f)}$\citet{1997A&A...322..924M}; $^{(g)}$\citet{2020A&A...641A..57M}.
}
\end{table*}

The fractional abundances of the molecules are parameterised according to the following expressions. For the molecular abundances, we adopted a profile \citep{2020A&A...641A..57M, Saberi19}
\begin{subequations}
\label{eq:ab_prof}
\begin{align}
f_{\mathrm{mol}} = f_{\mathrm{0}}\,\exp\{-(r/r_{\mathrm{p}})^{\alpha}\},
\end{align}
where $f_\mathrm{0}$ is the initial abundance at the starting radius of the profile, $r_\mathrm{p}$ is the characteristic photo-dissociation radius, and $\alpha$ is a parameter to model the steepness of the radial decrease around $r_\mathrm{p}$. In Table\,\ref{tab:parmol}, we present the abundance parameters that were used in our simulations.

We used the results of \citet{Saberi19} for the CO abundance profile. We note that they use an abundance profile 
\begin{align}
f_{\mathrm{mol}}^{\mathrm{Sab}} = f_{\mathrm{0}}\exp(-\ln(2) {(r/r_{1/2})^{\alpha}})= f_{\mathrm{0}}\,\left[\frac{1}{2}\right]^{(r/r_{1/2})^{\alpha}},
\end{align}
for their parameterization. We have adopted their values for the photo-dissociation radius to the profile we use (Eq.~\ref{eq:ab_prof}a), using $r_p = [\ln(2)]^{-1/\alpha} r_{1/2}$.

\end{subequations}
\begin{table}[hbtp!]
\caption{Molecular abundance profile parameters}
\label{tab:parmol}
\begin{center}
\begin{tabular}{l c c c}
\hline\hline
          & $f_{0}$ &  $r_{\mathrm{p}}$ (AU) & $\alpha$            \\
\hline              
CW Leo & & & \\
CO$^{a}$ & $6\times 10^{-4}$  & 19971 & 3.26 \\
H$^{13}$CN$^{b}$  & $5.6\times 10^{-7}$   & 2741 & 2  \\
SiS$^{c}$  & $1.3\times 10^{-6}$   & 1630 & 2  \\
CS$^{c}$  & $1.1\times 10^{-6}$   & 4593 & 2  \\
\hline
R Leo & & & \\
CO$^{a}$ & $2\times 10^{-4}$  & 1006 & 2.27 \\
H$^{13}$CN$^{b}$  & $4\times 10^{-8}$   & 401 & 2  \\
$^{29}$SiO$^{d}$ & $4.5\times 10^{-7}$   & 196 & 2  \\

\hline                                                      
\hline
\end{tabular}
\end{center}
\tablebib{
$^{(a)}$\citet{Saberi19}; $^{(b)}$\citet{Schoeier13} and \citet{2014A&A...566A.145R}; $^{(c)}$\citet{Massalkhi19}; $^{(d)}$\citet{2020A&A...641A..57M} and \citet{DeBeck2018}.
}
\end{table}

\subsection{Line polarisation simulations: CW~Leo}
\label{model_cwleo}
We modelled the polarised emission from the ALMA Band 7 transitions of the molecules CO, H$^{13}$CN, SiS and CS, towards a model of a circumstellar envelope, with parameters representing the CSE towards the AGB star CW Leo. 

The polarised emission of molecular lines is in the direction of the molecular alignment axis. In the interstellar medium, the alignment axis is generally with respect to the magnetic field direction, as precession around the magnetic field is fast relative to the other directional, radiative, interactions. However, in the CSEs that we investigate, the molecular lines of interest are excited in a region close to a strong stellar radiation source, thus inducing strong vibrational radiative interactions. It is possible that the stellar radiation field, that is typically in the radial direction of the CSE, determines the molecular alignment axis, if radiative transition rates exceed the magnetic precession rate. 

\begin{figure}[ht!]
\centering
\includegraphics[width=0.45\textwidth]{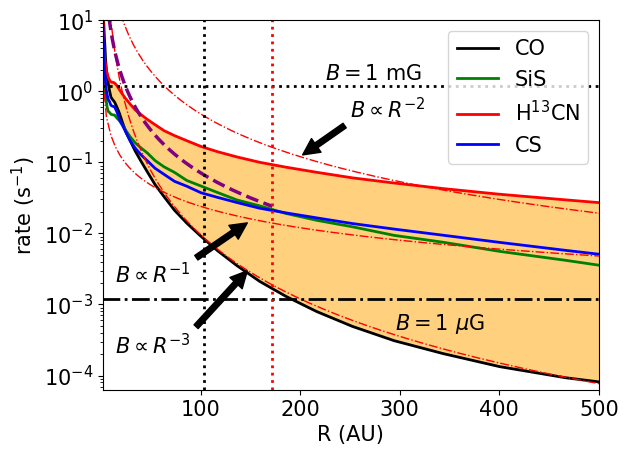}
\caption{Comparison of the radiative interaction (solid lines) and magnetic precession rates (dash-dotted lines) of the investigated molecules toward CW Leo. Magnetic precession rates are given for magnetic fields $B = B_{\mathrm{surf}} (R/R_{\mathrm{star}})^{-q}$, adopting $q=1$, $q=2$ and $3$, where for $q=1$ a $B_{\mathrm{surf}}=1$ mG is adopted, while for $q=2$ and $3$, $B_{\mathrm{surf}}=1$ G is adopted. The black long-dashed curve indicates the expected magnetic precession rates that explain the polarisation morphology of CS, CO and H$^{13}$CN within 1.4\arcsec. Horizontal black dotted and dash-dotted lines indicate the molecular magnetic precession rates at $1$ mG and $1$ $\mathrm{\mu} $G, respectively. Vertical black and red dotted line indicates telescope resolution and the 1.4\arcsec\, radius where a polarisation morphology change was observed (see \S~\ref{obs: CW Leo}). The orange coloured region indicate the magnetic field strengths, where CO would be aligned to the magnetic field, but H$^{13}$CN would be aligned in the radial direction, along the stellar radiation field.}
\label{fig:cw_rates}
\end{figure}

As a first step, to investigate the dominant axis of alignment, in Fig.~\ref{fig:cw_rates}, we compare for the upper levels of the investigated transitions, the radiative interaction rates to the magnetic precession rate, as a function of the distance to the central stellar object. The magnetic precession rate depends on the magnetic field strength. We give the magnetic precession rate assuming a surface magnetic field of $B_{\mathrm{surf}}=1$ G, and a $B \propto R^{-2}$ and a $B \propto R^{-3}$ distance relation, as well as $B_{\mathrm{surf}}=1$ mG with a $B \propto R^{-1}$ distance relation. Whilst the magnetic precession rate is unknown, it is striking that radiative interaction rates of the molecules CS, H$^{13}$CN and SiS far exceed the radiative interaction rates of CO. Indeed, assuming a magnetic field $B = 1\ \mathrm{G}\ (R/R_{*})^{-3}$, as would be typical for a dipole magnetic field, we predict that CO molecules align themselves to the magnetic field, whilst CS, H$^{13}$CN and SiS align to the (radially directed) radiation field. 

Concurrently, Figs.~\ref{IRCCO}-\ref{IRCSiS} show an intricate polarisation morphology for CO, consistent with non-radial alignment direction. The polarisation vectors in the CS, H$^{13}$CN and SiS observations tend to be tangential in regions beyond 1.4\,\arcsec, consistent with a radial alignment direction of the molecules. For offsets $<$1.4\,\arcsec, a transition into a different polarisation morphology seems to emerge for CS, and to a lesser degree, SiS. Strikingly, this polarisation morphology seems to agree with the morphology obtained from CO. We take this as an indication that CO is aligned with the magnetic field throughout the CSE, and that for $<$1.4\,\arcsec, the magnetic field is the dominant alignment direction also for CS and SiS. We find that H$^{13}$CN is aligned to the radiation field throughout the CSE, while CS and SiS are radiatively aligned for $R>$1.4\arcsec. The different alignment properties of the molecules allow us to estimate the magnetic field from the radiative interaction rates. For the entire (observed) CSE, we find that the magnetic field has to lie in the orange highlighted region in Fig.~\ref{fig:cw_rates}, whereas the different alignment properties for $R<$1.4\,\arcsec suggest magnetic field strengths $\gtrsim$ 20\,$\mathrm{ \mu G}$ $[R/(1.4 \arcsec)]^{-2} =$ 80\,$\mathrm{m G}$ $[R/R_*]^{-2}$, for $R<$1.4\,\arcsec. 

\begin{figure*}[ht!]
\centering
\begin{subfigure}[b]{0.33\textwidth}
\includegraphics[width=\textwidth]{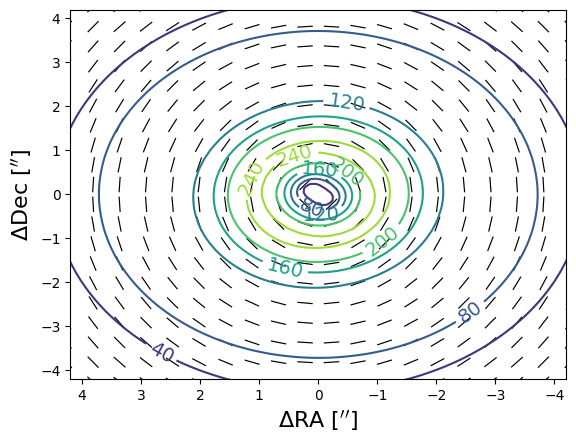}
\caption{H$^{13}$CN}
\end{subfigure}
\hfill
\begin{subfigure}[b]{0.33\textwidth}
\includegraphics[width=\textwidth]{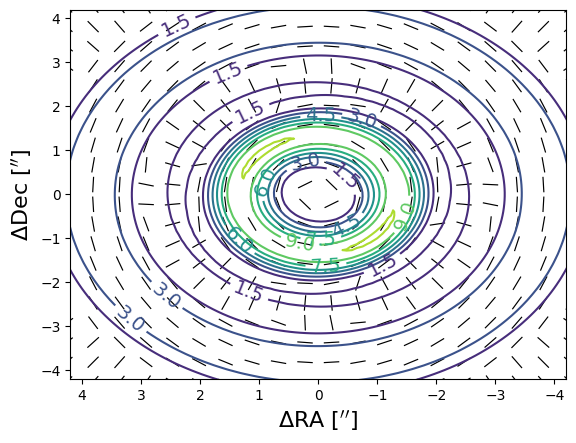}
\caption{CS}
\end{subfigure}
\hfill
\begin{subfigure}[b]{0.33\textwidth}
\includegraphics[width=\textwidth]{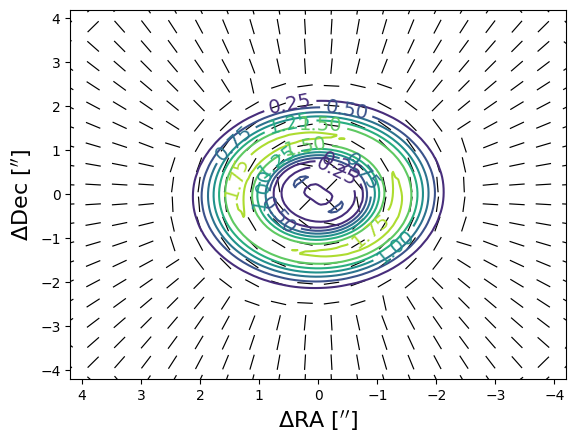}
\caption{SiS}
\end{subfigure}
\caption{Predicted polarised intensity contours (labelled in units of mJy/beam) overlaid with line segments that indicate the polarisation of (a) H$^{13}$CN, (b) CS, and (c) SiS emission emerging from a CSE representative of CW Leo. polarised intensity is given at the systemic velocity channel. The symmetry axis of the molecules is chosen in the radial direction, representative of radiative alignment. The simulated data has been convolved with the synthesised beam from the observations. The panels corresponds to the emission in a velocity channel centred on the stellar velocity.}
\label{fig:cw_rad}
\end{figure*}

In Fig.~\ref{fig:cw_rad}, we present PORTAL simulations of the transition lines of CS, H$^{13}$CN and SiS, excited towards CW Leo. For these simulations, we assumed the alignment axis to lie in the radial direction, which is the direction expected when molecules are radiatively aligned to the central stellar radiation field. Polarisation intensities and fractions are roughly consistent with the observations for CS and SiS. For H$^{13}$CN the PORTAL simulations predict a larger region where emission is polarised at a detectable level, as well as a azimuthally averaged polarisation fraction that peaks at $\sim4\%$. This is approximately two times higher than that seen in Fig.~\ref{Fig:Radial}. For H$^{13}$CN, we predict a tangential polarisation morphology throughout the emission region, which is consistent with observations. Likewise, for SiS, we predict a tangential polarisation morphology, which at offsets in excess of 1.4\,\arcsec~is consistent with the observations. The CS polarisation is tangential in the inner regions, but also shows a 90-degree flip to a radial morphology in a region with relatively low polarisation yields. For offsets $<$1.4\,\arcsec, we correctly predict the tangential polarisation morphology of H$^{13}$CN, while the predicted polarisation morphology of CS and SiS begins to diverge from the observed polarisation morphology. We suggest that this is an additional indication for a non-radial alignment in play, close to the central star.

\begin{figure*}[ht!]
\centering
\includegraphics[width=\textwidth]{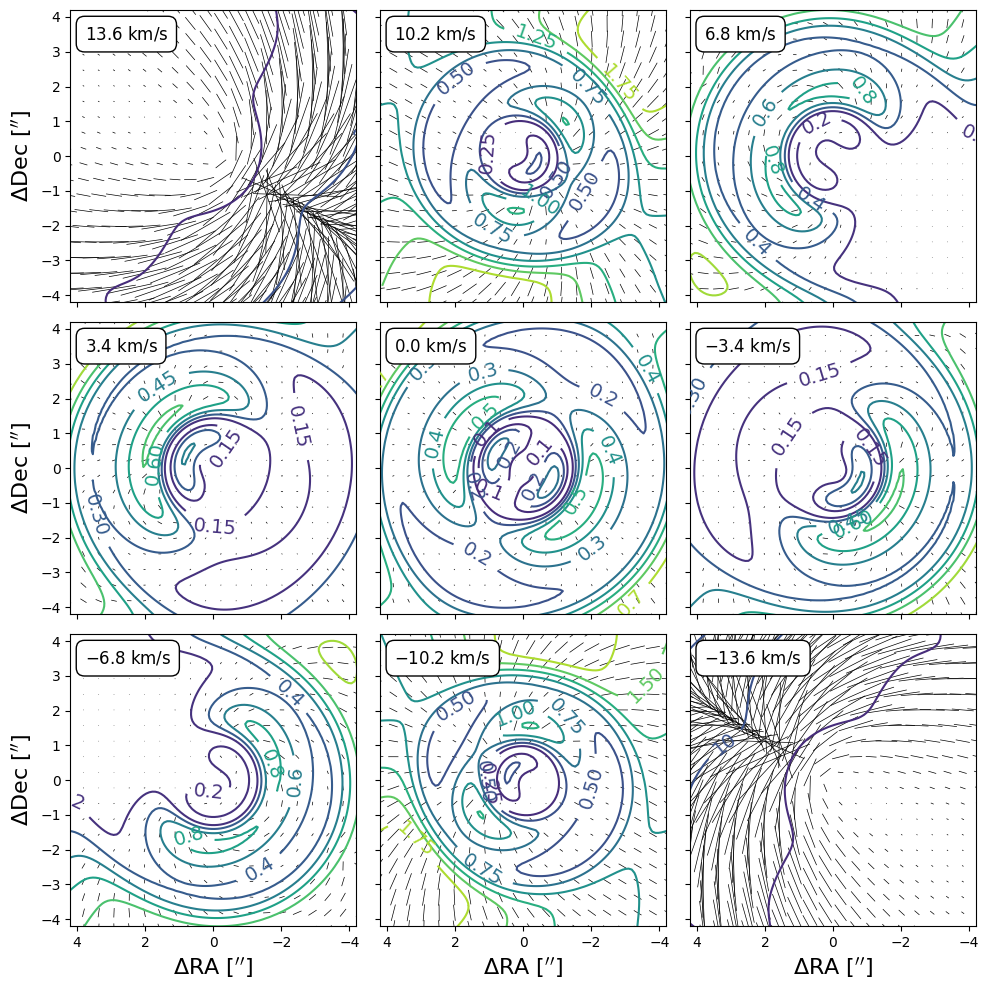}
\caption{As Fig.~\ref{fig:cw_rad} for a range of velocity channels of CO emission emerging from a CSE representative of CW Leo. The symmetry axis of the molecules is chosen to be along the magnetic field, which is assumed as a dipole field with position angle $=125\degr$ and inclination $45\degr$. For this figure, the line segments are scaled according to the polarisation fraction.}
\label{fig:cw_CO}
\end{figure*}

Fig.~\ref{fig:cw_CO}, we present PORTAL simulations of CO emission excited towards CW Leo. For these simulations, we assumed the alignment axis to be determined by a dipole magnetic field. We assume the dipole to be directed with a position angle of 125\degr, as was suggested by \citet{Andersson23}, and we assume it to be inclined at 45\degr. Towards the systemic velocity, we predict the polarisation morphology to adhere to a dipolar morphology, while at larger velocity offsets, the polarisation geometry becomes more intricate, with pronounced asymmetry between either sides of the dipole position angle. In addition, larger polarisation degrees are expected at large velocity offsets. Comparing the predicted polarisation morphology to the observed polarisation morphology in CO, we find only modest agreement. In Fig.~(\ref{fig:cw_CO_tor}) of the appendix, we furthermore investigate the polarisation signature of a toroidal magnetic field in the spectral lines of CO. Similar to the dipolar configuration, the symmetry axis is inclined at $45^o$, while the position angle is put at $125^o$. As was the case for the dipolar polarisation morphology, only modest agreement with the observations is found. This likely indicates that the morphology of the magnetic field in the envelope of CW~Leo cannot be described by a simple magnetic field geometry.

\subsection{Line polarisation simulations: R~Leo}
\label{model_rleo}
We modelled the polarised emission from the ALMA Band 7 transitions of the molecules CO, H$^{13}$CN and $^{29}$SiO, towards a model of a circumstellar envelope, with parameters representing the CSE towards the AGB star R Leo. 

\begin{figure}[ht!]
\centering
\includegraphics[width=0.45\textwidth]{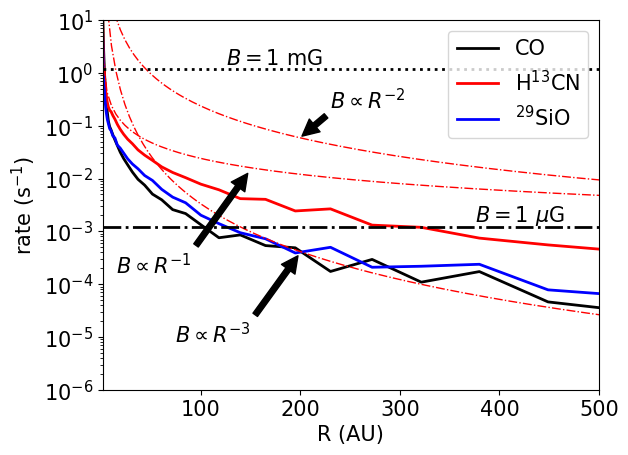}
\caption{Comparison of the radiative interaction (solid lines) and magnetic precession rates (dash-dotted lines) of the investigated molecules toward R Leo. Magnetic precession rates are given for magnetic fields $B = B_{\mathrm{surf}} (R/R_{\mathrm{star}})^{-q}$, adopting $q=2$ and $3$. Black dotted and dash-dotted lines indicate the molecular magnetic precession rates at $1$ mG and $1$ $\mathrm{\mu} $G, respectively.}
\label{fig:r_rates}
\end{figure}

As was done for CW Leo, as a first step, to investigate the dominant axis of alignment, in Fig.~\ref{fig:r_rates}, we compare for the upper levels of the investigated transitions, the radiative interaction rates to the magnetic precession rate, as a function of the distance to the central stellar object. The magnetic precession rate depends on the magnetic field strength. Magnetic precession rates are given for magnetic fields $B = B_{\mathrm{surf}} (R/R_{\mathrm{star}})^{-q}$, adopting $q=1$, $q=2$ and $3$, where for $q=1$ a $B_{\mathrm{surf}}=1$ mG is adopted, while for $q=2$ and $3$, $B_{\mathrm{surf}}=1$ G is adopted. We find low rates of radiative interactions for all investigated molecules. We find that beyond 100\,AU, a magnetic field in excess of 10 $\mathrm{\mu G}$ would determine the alignment direction.

\begin{figure*}[ht!]
\centering
\begin{subfigure}[b]{0.33\textwidth}
\includegraphics[width=\textwidth]{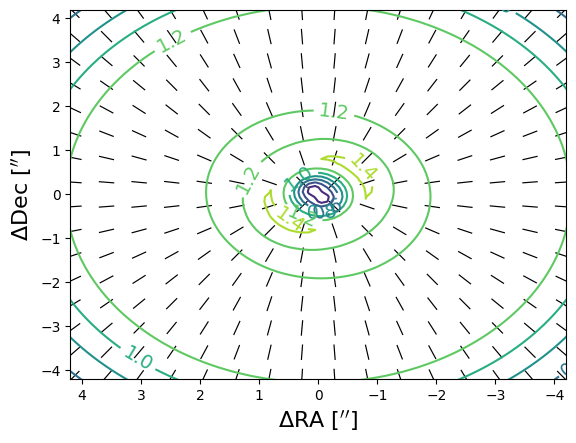}
\caption{CO}
\end{subfigure}
\hfill
\begin{subfigure}[b]{0.33\textwidth}
\includegraphics[width=\textwidth]{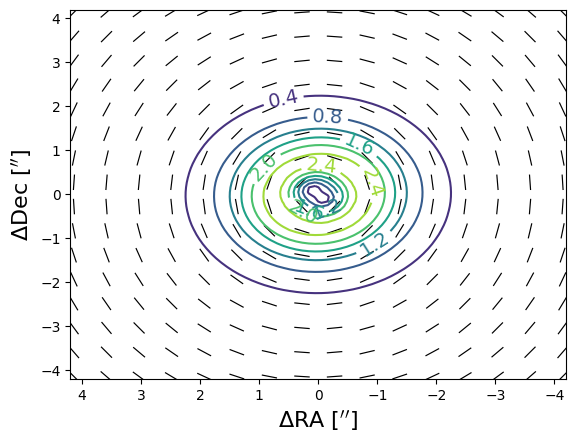}
\caption{H$^{13}$CN}
\end{subfigure}
\hfill
\begin{subfigure}[b]{0.33\textwidth}
\includegraphics[width=\textwidth]{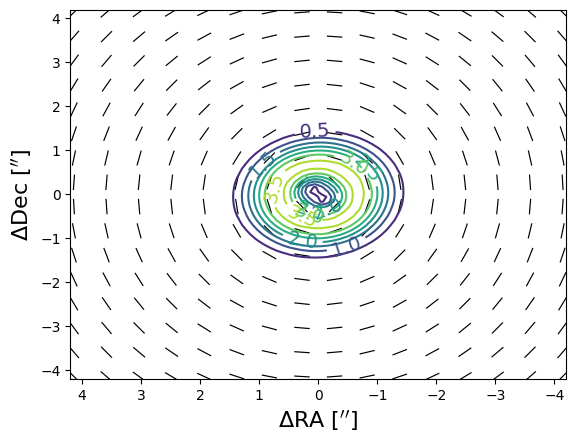}
\caption{$^{29}$SiO}
\end{subfigure}
\caption{As Fig.~\ref{fig:cw_rad} for the (a) CO, (b) H$^{13}$CN, (c) $^{29}$SiO emission emerging from a CSE representative of R Leo. polarised intensity is given at the systemic velocity channel. The symmetry axis of the molecules is chosen in the radial direction.}
\label{fig:r_rad}
\end{figure*}

Due to the low radiative interaction rates, we expect that the magnetic field determines the symmetry axes of the investigated molecules. Thus, the polarisation morphologies seen towards R Leo and reported in Figs.~\ref{RLeoCO}-\ref{RLeo29SiO}, suggest a radial magnetic field, which is likely the result of the influence of advection due to the outflow. In Fig.~\ref{fig:r_rad}, we present PORTAL simulations of the transition lines of CO, H$^{13}$CN and $^{29}$SiO, excited towards R Leo, assuming a radial symmetry axis. Significant ($>0.5\%$) polarisation is found for CO in regions within $2.5$\arcsec, while for H$^{13}$CN and $^{29}$SiO significant polarisation is found within $1.5$\arcsec.

The predicted polarisation morphology appears to be consistent with the observations. Crucially, as in the observations, for CO, we predict a radial polarisation morphology, while H$^{13}$CN and $^{29}$SiO the polarisation morphology is predicted to be tangential. Also, the regions with significant polarisation agree with the observations. In the observations, the polarisation degree of CO throughout the region varies strongly, which we do not find in our simulations. It should be noted, though, that the total intensity shows a variable emission pattern as well, which is likely due to an inhomogeneous outflow. Such clumpy structures may locally introduce an additional radiation anisotropy, which results in local enhancements of the polarisation degrees. Indeed, the regions showing intensity fluctuations in the CO observations of R Leo seem to be associated with high degrees of polarisation.

\section{Discussion}
\label{disc}
\subsection{Comparison with previous observations}

The molecular line polarisation arising from the GK-effect has previously been observed for both CW~Leo and R~Leo with, respectively, the SMA \citep{Girart12} and CARMA \citep{Huang20}. The SMA CW~Leo observations targeted the same molecular transitions and detected polarisation up to a level of $\sim6\sigma$ at a spatial resolution of $\sim1.5\times3$\arcsec and using a spectral resolution of $20$~km~s$^{-1}$ for CO $J=3-2$, CS $J=7-6$, and SiS $J=19-18$. The very different spatial and spectral resolution as well as limited sensitivity of the SMA observations makes a direct comparison difficult. The peak polarisation levels in the SMA observations were of the order of $\sim2\%$ for CO and SiS, and $\sim4\%$ for the CS. The SiS (in the right panel of figure 2 in \citet{Girart12} but wrongly identified as the middle panel in the caption) shows the same tangential morphology as seen in our observations. The polarisation of the CS (in the middle panel of figure 2 in \citet{Girart12} but wrongly identified as the right panel) is however stronger in the SMA observations and located in a region outside of the area where the polarisation is detected in the ALMA observations. CO polarisation was limited to a single, large, beam in the SMA observations and cannot be compared directly even if the fractional polarisation is similar. The differences between the SMA and ALMA observations can likely be attributed to the relatively low significance of the polarisation signal in the SMA observations as well as the differences in resolution and spatial filtering of the emission.

The CARMA observations of R~Leo \citep{Huang20} included the CO $J=2-1$ and SiO $v=1, J=5-4$ maser line at a spatial resolution of $\sim4\times5$\arcsec and a spectral resolution of 1~km~s$^{-1}$. The CO polarisation was detected at a level of $\sim6\sigma$ and a high fractional polarisation level of $\sim10\%$. The fractional polarisation is expected to be higher at the lower-$J$ transitions of CO, but considering the low significance and difference in spatial filtering we cannot make a relevant comparison. The polarisation vector direction shown in \citet{Huang20} is consistent with a radial polarisation morphology. The SiO maser was detected with a fractional polarisation of $\sim35\%$, similar to the level measured in our observations for the $J=8-7$ transition. Although the direction of the polarisation reported in table 3 of \citet{Huang20} for the SiO maser is different, the direction seen in their figure 5 is fully consistent with the direction shown in our Fig.~\ref{RLeoSiOv1}. The good correspondence between the direction of polarisation of the two different maser transitions further supports the existence of a preferred magnetic field direction in the SiO maser region close to the star. Observations of the Zeeman splitting by \citet{Herpin06} of SiO $v=1, J=2-1$ indicate a field strength in the SiO maser region between $B=4-5$~G. 

Continuum polarisation has also been observed in the extended dust envelope around CW~Leo using SOFIA \citep{Andersson22} and SCUBA-2 \citep{Andersson23}. The ALMA observations cover much less than a single, central, beam of the JCTM SCUBA-2 observations. We adopted the dipole model described in \citet{Andersson23} in our model shown in Fig.~\ref{fig:cw_CO}, but, because of the large difference in scales, can not make more than a qualitative comparison. It appears that, at the scales probed by ALMA, the magnetic field morphology is more complex than a dipole field. Our continuum polarisation direction, as well as the dominant direction of the CO polarisation towards the star, is consistent with the single SCUBA-2 vector in the region probed by the ALMA observations. Unfortunately, spatial interferometric filtering and the low surface brightness in the extended envelope at ALMA resolution, does not allow us to compare the extended emission between SCUBA-2 (and SOFIA) and ALMA even with mosaiced observations.  

\subsection{Anisotropic Resonant Scattering}

In addition to the generation of linear polarisation due to the GK-effect when emission passes through a magnetised molecular region, linear polarisation can also be converted into circular polarisation in an effect known as anisotropic resonant scattering \citep{Houde13, Houde22}. As a result, the linear polarisation might no longer trace the magnetic field direction. 

Since the circular polarisation of the extended lines in our observations is affected by the beam squint of the ALMA antennas \citep[see also][]{Teague21}, we cannot directly investigate anisotropic resonance scattering through a study of the circular polarisation. As argued in \citet{Vlemmings23}, the possible anisotropic scattering of molecular lines in a circumstellar envelope is likely limited to that arising from the emitting region itself, since the limits imposed by photo-dissociation and large velocity gradients severely reduces the available molecular column density of the foreground. In \citet{Chamma18}, the SMA polarisation observations of CW~Leo were analysed, and it was concluded that significant anisotropic resonant scattering was present in all of the lines also observed with ALMA. This was based on the detection of several percent of circular polarisation in different regions of the circumstellar envelope. However, this would result in significant rotation (and in some cases complete depolarisation) of the linear polarisation vectors, with strong variation across the envelope. This would have ruled out the detection of the structured radial or tangential polarisation morphology we observe. But we detect no sign of this behaviour in our linear polarisation maps of, in particular, the CS, SiS and H$^{13}$CN. This leads us to conclude that the circular polarisation detected towards CW~Leo in \citet{Chamma18} is not the results of anisotropic resonant scattering, but rather an instrumental effect. Since we do not appear to see the signatures of vector rotation towards most of the lines, we conclude that anisotropic resonant scattering likely does not affect the polarisation of our sources. A firm conclusion can only be reached when high quality circular polarisation observations become available.

\begin{figure}[ht!]
         \centering
         \includegraphics[width=0.45\textwidth]{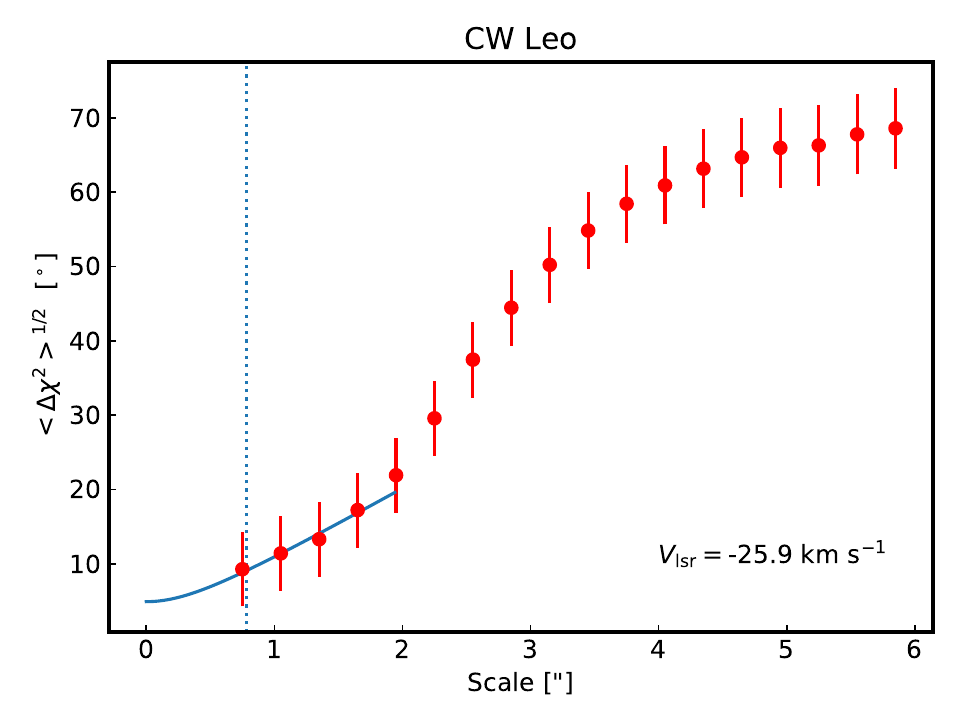}
        \caption{Dispersion of the polarisation vectors (the square root of the second order structure function), binned to the Nyquist sampled resolution, for the central velocity channel of the CO $J=3-2$ polarised emission around CW~Leo. The error bars indicate the variance in each bin. The vertical dotted line indicates the size of the beam major axis. The solid line indicates the fit of the Structure Function Analysis described in the text.}
        \label{fig:SFACWLeo}
\end{figure}

\subsection{Magnetic field strength: Structure Function Analysis}
\label{SFA}

Assuming that the polarisation vectors of the CO $J=3-2$ transition correspond to the large scale magnetic field around CW~Leo, we can use a Structure Function Analysis (SFA), often used for magnetic field studies of star forming regions, to calculate the ratio between turbulent and mean large scale magnetic field strength \citep[e.g.][]{Hildebrand09, Houde09}. 

In our analysis, we follow the equations from \citet{Koch10} and the steps described in \citet{Vlemmings23}.
Under the assumption that the turbulent field arises from transverse Alfv{\'e}n waves in an environment with isotropic and incompressible turbulence in which the magnetic field is frozen into the gas, the ratio between the turbulent ($B_t$) and large scale magnetic field component  $B_0$ is equal to the ratio between the turbulent line width $\sigma_\nu$ and the Alfv{\'e}n velocity $\sigma_A=\frac{B_0}{\sqrt{4\pi\rho}}$. Here $\rho$ is the density of the gas. Assuming an average density and turbulent velocity of the emitting region of the CO gas, we can then derive an estimate for the strength of the plane of the sky component of large scale magnetic field. In the case that the magnetic field is not fully frozen in the gas, which is, considering the observed morphology of the magnetic field, likely the case around CW~Leo, the derived field strength should be considered a lower limit. 

Fig.~\ref{fig:SFACWLeo} shows the result of the SFA for the central velocity channel of the CO emission around CW~Leo. The dispersion of the polarisation vectors is shown to steadily increase from the small scales until asymptotically approaching a plateau of $\sim70^\circ$ in a behaviour of the structure function similar to that observed for several star forming regions \citep[e.g.][]{DallOlio19}. The structure function can be fit using the equation from \citet{Koch10} within $\lesssim2$\arcsec, which is the characteristic length scale for variations in the large scale magnetic field component. We find that the ratio between the turbulent and large scale magnetic field component $\frac{\langle B_t^2\rangle^{1/2}}{B_0} = 0.061\pm0.001$. Similar to \citet{Vlemmings23}, this means that we can write the strength of the large scale magnetic field strength as: 
\begin{equation}
    B_0=19~\Biggl( \frac{\langle n_{H_2}\rangle}{10^6}\Biggr)^{1/2} \frac{v_t}{1.5}~{\rm mG},
    \label{Eq:B}
\end{equation}
 with the average H$_2$ number density, $n_{H_2}$, in the CO region in units of cm$^{-3}$ and the typical turbulent velocity, $v_t$, in km~s$^{-1}$. The adopted density corresponds to a radius of $\sim250$~au. Using Eq.\ref{Eq:n} and the CW~Leo mass-loss rate and expansion velocity, we can also describe the magnetic field as a function of radius in the region of the ALMA observations (between the Nyquist sampled beam and maximum recoverable scale) as $B=107\times(50/r)$~mG between $r=50$ and $485$~au. 
 The error on the derived field strength is completely dominated by the uncertainty in the average number density and the assumption of magnetic flux freezing. A further source of uncertainty is the effect of limited angular resolution and spatial filtering \citep{Houde16}. Following the approach outlined in \citet{Vlemmings23}, the number of independent turbulent cells ($N$) probed by our observations is estimated using the formula from \citet{Houde16}:
\begin{equation}
    N = \frac{(\delta^2 + 2W_1^2)\Delta'}{\sqrt{2\pi}\delta^2}.
\end{equation}
We use the approximation that the correlation length, $\delta$, is set by the typical clump size $r_c\propto r^{0.8}$, while the depth of the observed molecular layer along the line of sight, $\Delta'$, is estimated from the total size of the CO envelope divided by the number of channels \citep[for a derivation of both of these, see][]{Vlemmings23}. This yields $\delta\sim300$~au and $\Delta'\sim500$~au. The parameter $W_1\approx40$~au is the radius of the interferometric beam. Hence, we find $N\sim1.2$. According to \citet{Houde16}, the ratio between the turbulent and large scale magnetic field strengths scales with $\sqrt{N}$. This means that our derived magnetic field strength does not require a significant correction.

 An analysis of the other velocity channels for which sufficient independent polarisation vectors where measured yield similar field strengths to within $30\%$. The field strength is $2-10$ times higher than the line-of-sight magnetic field strength of $|B_{||}|\sim2-9$~mG estimated from Zeeman splitting of CN in the outer part ($R\sim2500$~au) of the envelope of CW~Leo \citep{Duthu17}. Since we measure the (plane-of-the-sky) field component in CO much closer to the star ($R\sim 250$~au), the comparison between the CO estimates and possible CN Zeeman measurements implies that the magnetic field strength beyond $R\sim250$~au decreases outward as $R^{-q}$ with $0.3<q<1.0$. Taking into account the large uncertainties, this is only marginally consistent with a dominating toroidal magnetic field component at these radii ($q=1$). A magnetic field frozen into the spherically expanding gas or a solar-type magnetic field ($q=2$), as well as a dipole-shaped or radial magnetic field ($q=3$) appear to be ruled out. 
 
\begin{figure}[t!]
         \centering
         \includegraphics[width=0.45\textwidth]{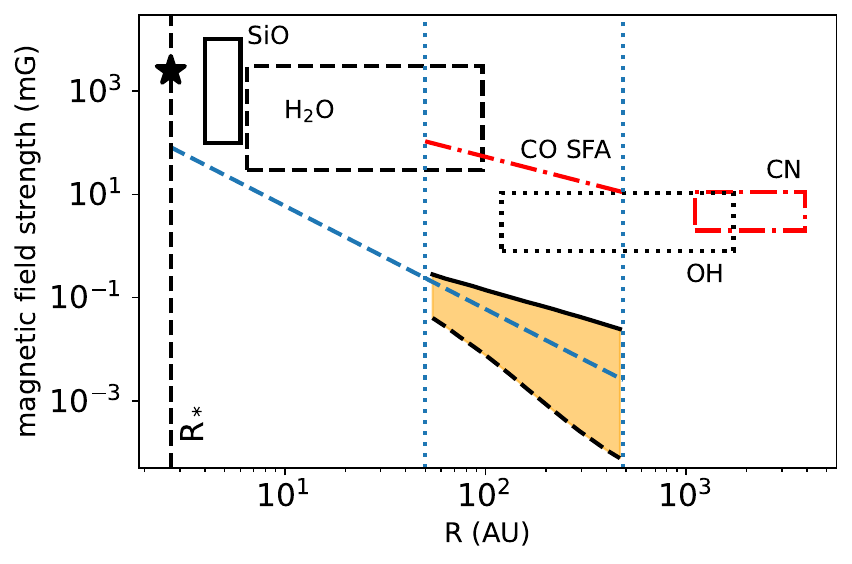}
        \caption{Magnetic field strength in AGB circumstellar envelopes as a function of radial distance. The black boxes indicate the literature values reported from maser measurements \citep[e.g.][]{Herpin06, LealFerreira13, Vlemmings05, Rudnitski10}. The black star indicates the surface magnetic field measurement for the star $\chi$~Cyg \citep{Lebre14}. The measurements fro CW~Leo are indicated in red. These are the CN Zeeman measurements from \citet{Duthu17} (red dash-dotted box) and the CO structure function analysis result reported in \S~\ref{SFA} (red dashed-dotted line). For the CO result we use the relation between $n_{\rm {H}_2}$ and $R$ for the range of radii probed by the ALMA observations (Eq.~\ref{Eq:B}). This range is indicated by the vertical dotted lines. The stellar surface is indicated by the long-dashed vertical line. We indicate, in orange, for the region probed by ALMA around CW~Leo, the lower (dashed) and upper (solid) magnetic field limits based on the analysis of the $^{12}$CO and H$^{13}$CN radiative interaction rates. The relation between radius and magnetic field strength that fits the observations and modelled radiative rates described in \S~\ref{model_cwleo} is presented as the blue dashed relation (extrapolated out the the maximum scale probed with the ALMA polarisation observations). It is clear that this relation is not in agreement with the other magnetic field strength observations.}
        \label{fig:BvsR}
\end{figure}

\subsection{Magnetic field strength: molecular alignment}

As described in \S~\ref{sec:polmods} and shown in Figs.~\ref{fig:cw_rates} and \ref{fig:r_rates}, a comparison between the radiative excitation and magnetic precession rates for various magnetic tracers throughout the circumstellar envelope can also be used to derive an estimate of the magnetic field strength. 

We present such a comparison in Fig.~\ref{fig:BvsR}.
For the relatively low-density outflow around R~Leo, it is clear that the magnetic precession rate dominates the radiative rates of our observed molecular lines throughout the entire envelope, even for a very low ($\sim10~\mu$G) magnetic field strength. As described in \S~\ref{model_rleo}, a radial magnetic field can reproduce all observed lines. This would imply that, around R~Leo, the magnetic field in much of the envelope is coupled to the outflow, but that the kinetic outflow energy dominates the magnetic energy resulting in the magnetic field lines being stretched by the predominantly radial outflow.

For the much denser envelope of CW~Leo, the different behaviour of the various molecular tracers places constraints on the magnetic field strength throughout the envelope. However, these constraints are not consistent with both the magnetic field strength determined from the structure function analysis of the CO emission as well as the CN Zeeman measurements in the outer envelope \citep{Duthu17}. Specifically, the fact that, beyond $\sim1.4\arcsec$, the CS, SiS and H$^{13}$CN all apparently trace the direction imposed by the radiation field, while the CO still traces the magnetic field, seems to rule out a magnetic field strength $\gtrsim 20\mu$G at $R\sim170$~au and beyond. As discussed previously, the CN Zeeman observations appear to indicate $|B_{||}|\sim2-9$~mG at $\sim2500$~au and the CO structure function analysis implies $|B_{\perp}|\sim19$~mG at $\sim250$~au. Under the conditions of such a magnetic field, also CS, SiS and H$^{13}$CN should be aligned with the magnetic field in the outer envelope.
One possible solution to this discrepancy is if our molecular analysis did not include all relevant molecular transitions through which the molecular alignment can occur. Should radiative molecular alignment for the species we observe be more effective in for example higher vibrational transitions that were not considered, the radiative alignment rates in our models would increase. This would allow for a stronger magnetic field while maintaining the radiative alignment of CS, SiS and H$^{13}$CN in the outer envelope. 

\section{Conclusions}
\label{conc}
We present a detailed analysis of the molecular line polarisation observed with ALMA in the circumstellar envelopes of the C-type AGB star CW~Leo and the M-type AGB star R~Leo. In both sources, we detect line polarisation of multiple molecular lines, with fractions ranging from $1.18\%$ for $^{12}$CO to $32.4\%$ for one of the masing SiO lines. We also detect polarisation of the continuum, but since most dust continuum emission is filtered out by the interferometric observations, this is limited to a single beam towards the stars. Around CW~Leo, the CO polarisation likely traces a complex large scale magnetic field. Within $R=1.4\arcsec\approx170$~au, CS traces the same structure. The polarisation vectors of SiS and H$^{13}$CN, as well as those of CS beyond 170~au, are predominantly tangential, indicating molecular alignment due to the radiation field. A structure function analysis of the CO, reveal a plane-of-the-sky magnetic field of $\sim19$~mG at $R=250$~au, which would be consistent with the values obtained from CN Zeeman observations \citep{Duthu17}. These values however, are too high to explain the alignment of CS, SiS and H$^{13}$CN with the radiation field in the outer envelope. Additional modelling and observations are needed to solve this discrepancy. Around R~Leo, the observed polarisation morphology of the CO, H$^{13}$CN and $^{29}$SiO can be explained by a large scale magnetic field that is radially advected by the outflow. The observations and modelling presented here show that polarisation observations of multiple molecular species can be a powerful tool to determine both the magnetic field but also the behaviour of the radiation field throughout circumstellar envelopes.

\begin{acknowledgements}
BL acknowledges VR support under grant No. 2021-00339. This paper makes use of the following ALMA data: ADS/JAO.ALMA\#2016.1.00251.S. ALMA is a partnership of ESO
  (representing its member states), NSF (USA) and NINS (Japan),
  together with NRC (Canada), NSC and ASIAA (Taiwan), and KASI
  (Republic of Korea), in cooperation with the Republic of Chile. The
  Joint ALMA Observatory is operated by ESO, AUI/NRAO and NAOJ. The project leading to this publication has received support from
ORP, that is funded by the European Union’s Horizon 2020 research
and innovation programme under grant agreement No 101004719 [ORP]. 
LV-P acknowledges support from the
grant PID2020-117034RJ-I00 funded by the Spanish MCIN/AEI/ 10.13039/501100011033. The computations were enabled by resources provided by Chalmers e-Commons at Chalmers.
We also acknowledge support from the Nordic ALMA Regional Centre (ARC)
  node based at Onsala Space Observatory. The Nordic ARC node is
\end{acknowledgements}

%
%

\bibliographystyle{aa}

\begin{appendix}
\section{Maser transitions}
\begin{figure*}[ht!]
\centering
\includegraphics[width=0.95\textwidth]{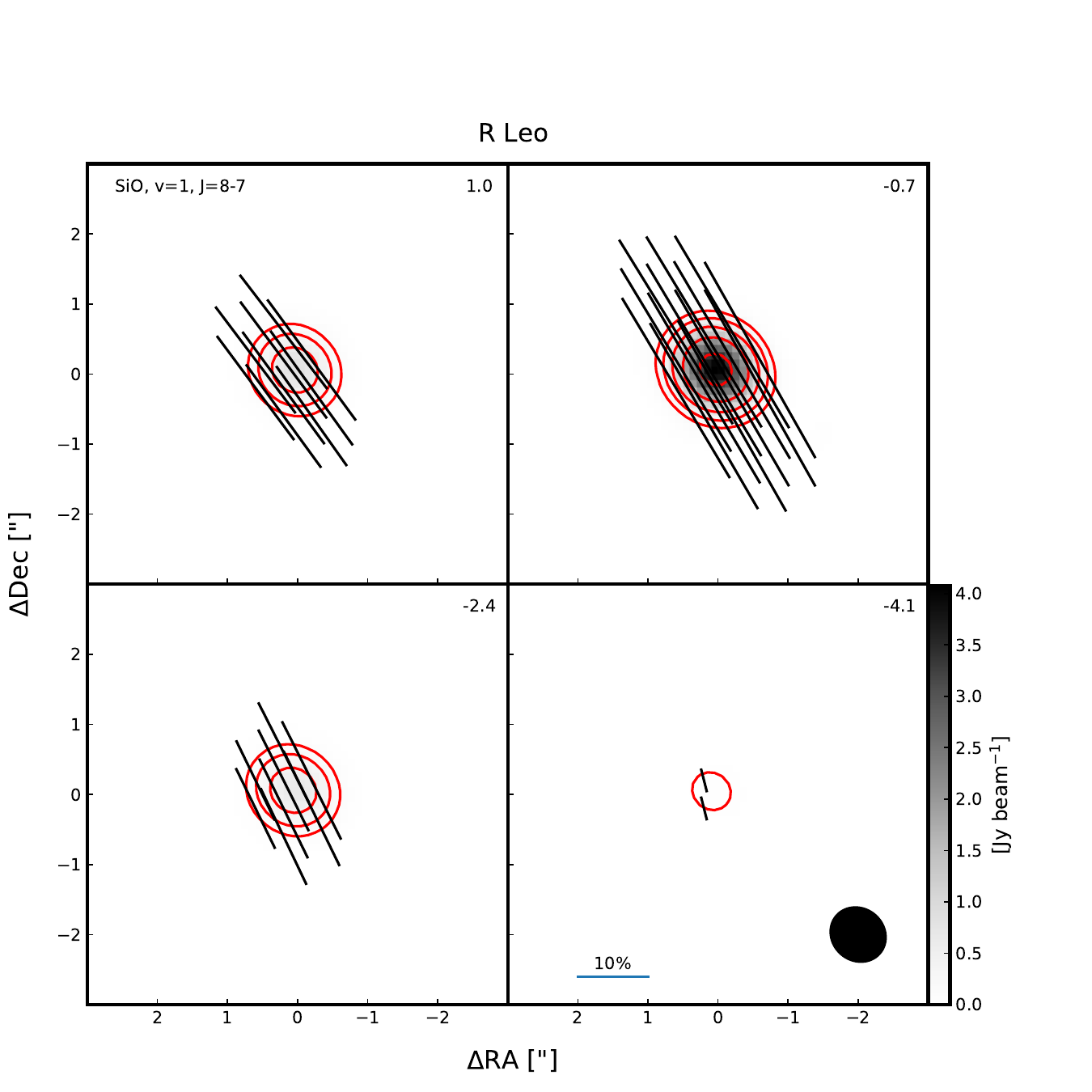}
\caption{Same as Fig.~\ref{IRCCO} for the SiO $v=1, J=8-7$ maser emission around R~Leo. The peak emission is $I_{\rm SiOv1, peak}$=12.86~Jy~beam$^{-1}$. The maximum polarisation $P_{\rm l, max}= 32.4\%$.}
\label{RLeoSiOv1}
\end{figure*}

\begin{figure*}[ht!]
\centering
\includegraphics[width=0.95\textwidth]{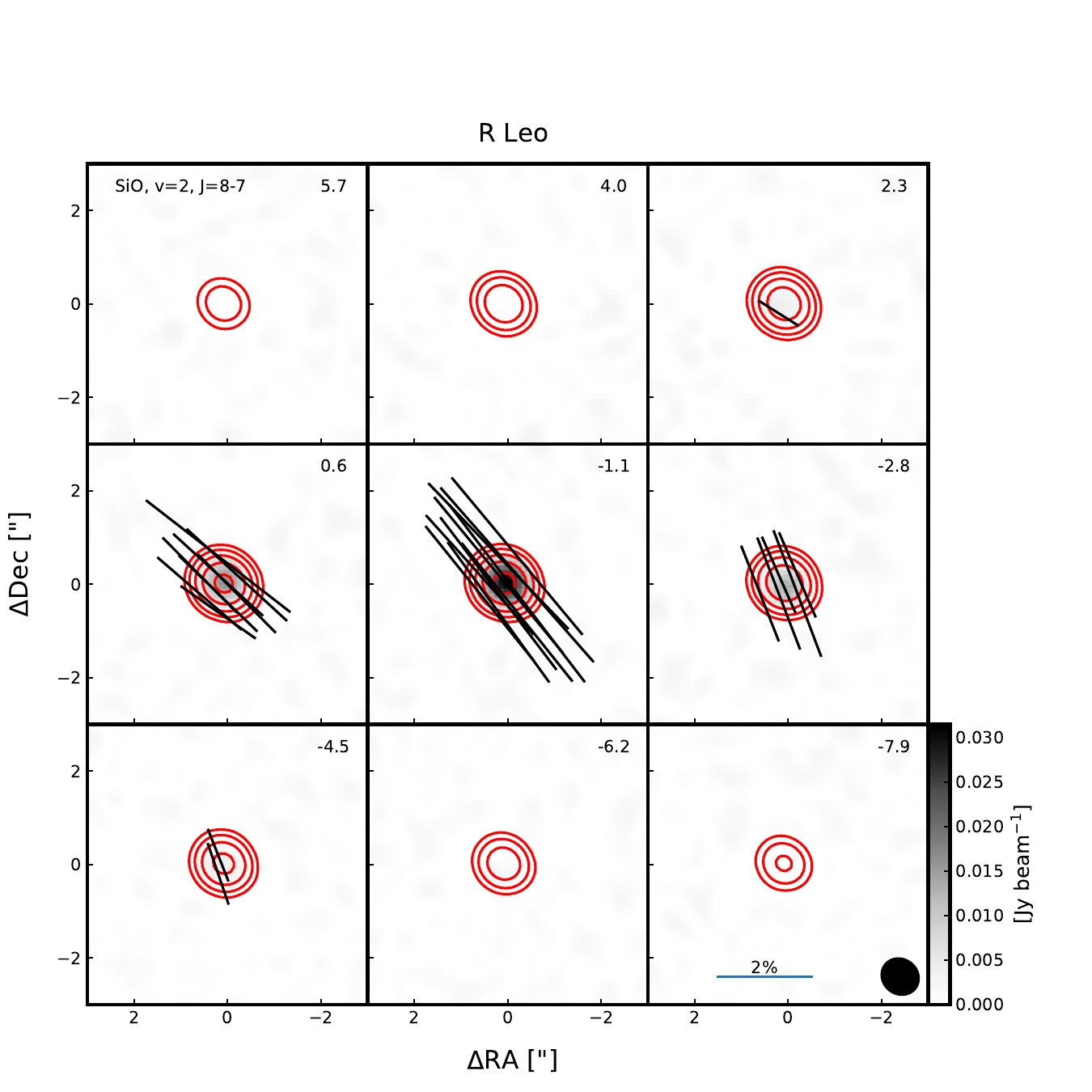}
\caption{Same as Fig.~\ref{IRCCO} for the SiO $v=2, J=8-7$ emission (which likely includes maser amplification) around R~Leo. The peak emission is $I_{\rm SiOv1, peak}$=0.72~Jy~beam$^{-1}$. The maximum polarisation $P_{\rm l, max}= 4.92\%$.}
\label{RLeoSiOv2}
\end{figure*}

In Figs.~\ref{RLeoSiOv1} and \ref{RLeoSiOv2} we present the compact, polarised, SiO maser emission observed around R~Leo.

\section{polarisation signature of a toroidal magnetic field}

\begin{figure*}[ht!]
\centering
\includegraphics[width=\textwidth]{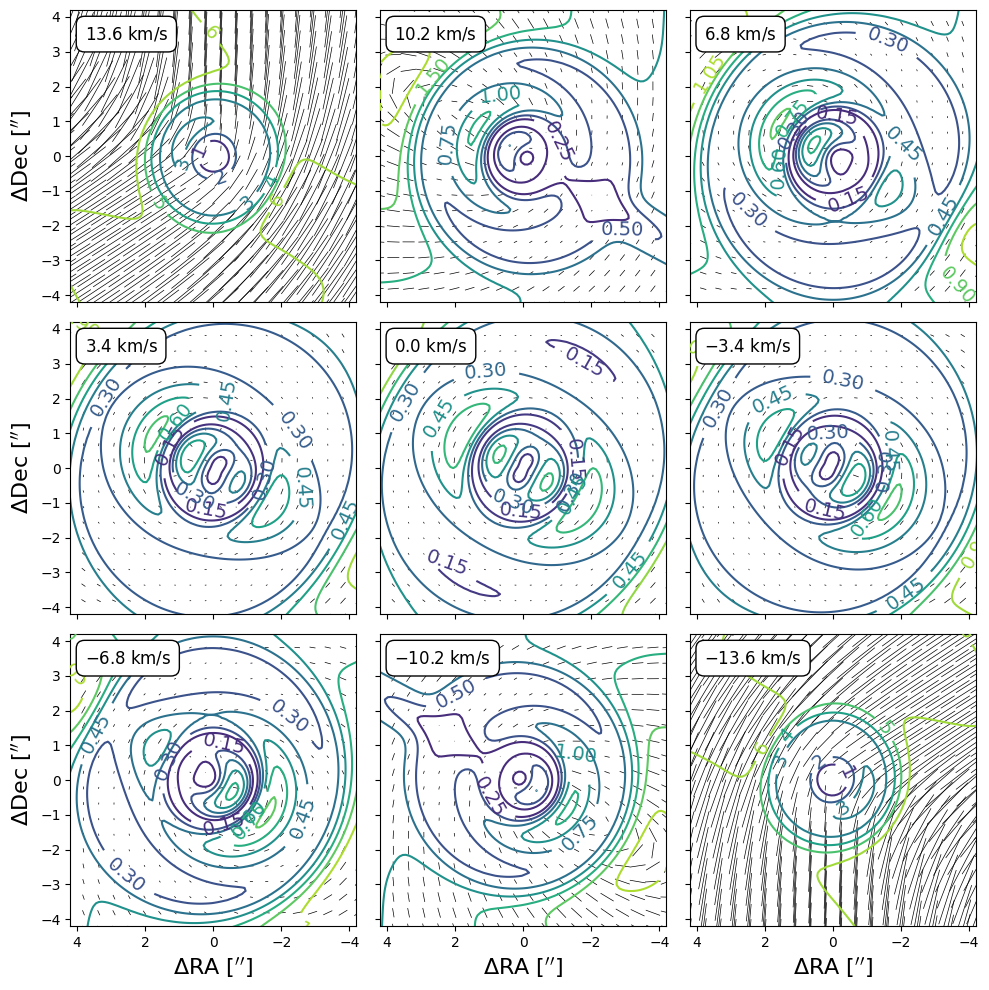}
\caption{As Fig.~\ref{fig:cw_CO} for a range of velocity channels of CO emission emerging from a CSE representative of CW Leo. Here, the symmetry axis of the molecules is chosen to be along the magnetic field, which is assumed as a toroidal field with position angle $=125\degr$ and inclination $45\degr$. Simulated data has been convolved with the synthesised beam from the observations. The panels are labelled with the velocity with respect to the stellar velocity.}
\label{fig:cw_CO_tor}
\end{figure*}

In Fig.~\ref{fig:cw_CO_tor} we present the model for the CO polarisation in the circumstellar envelope of CW~Leo when assuming a purely toroidal magnetic field configuration at the same position angle and inclination as the dipole field model. Note that a purely toroidal field is likely nonphysical in a circumstellar envelope, but the model represents the theoretical extreme of a magnetic field with a dominant toroidal component.

\end{appendix}

\end{document}